\pdfoutput=1
\documentclass[twoside,slac_one]{revtex4}

\usepackage{graphicx}
\usepackage{fancyhdr}
\usepackage{amsmath} % American Mathematics Society standards
\usepackage{bm}% bold math
\usepackage{amsxtra}
\usepackage{amssymb}
\usepackage{amsthm}
\usepackage{latexsym}
\usepackage{lscape}

\newcommand{\Bs}{\overline{B}_s^0}
\newcommand{\thickhline}{\noalign{\hrule height 0.8pt}}
\def\etal{{\it et al}}
\usepackage{afterpage}

\begin{document}

%Title of paper
\title{Heavy Flavor Physics}

% Repeat the \author .. \affiliation  etc. as needed
%
% \affiliation command applies to all authors since the last
% \affiliation command. The \affiliation command should follow the
% other information

\author{Sheldon Stone}
\affiliation{Department of Physics, Syracuse University, Syracuse, NY, USA and CERN, Geneve, SW}

\begin{abstract}
The main purpose of Heavy Flavor experiments is to discover physics beyond the Standard Model, or characterize it, should it be found elsewhere. Thus, current limits on New Physics (NP) are reviewed.  New results are presented, some involving processes that could show NP  even with current data. Specific topics include the CKM element $|V_{ub}|$, the forward-backward asymmetry in $\overline{B}^0\to \overline{K}^{*0}\mu^+\mu^-$, $b$-hadron fractions at the LHC, ${\cal{B}}$($\Bs\to\mu^+\mu^-$), first observations of several $\Bs$ and $B_c^-$ decay modes, the $X(4140)$, new $b$-baryons and their decays, searches for Majorana neutrinos, and Lepton Flavor Violation.
\end{abstract}

%\maketitle must follow title, authors, abstract
\maketitle

\thispagestyle{fancy}

% body of paper here - Use proper section commands
% References should be done using the \cite, \ref, and \label commands
% Put \label in argument of \section for cross-referencing
%\section{\label{}}

%%%%%%%%%%%%%%%%%%%%%%%%%%%%%%%%%%
\section{Introduction}
Heavy Flavor Physics (HFP) is the study of interactions that differ among flavors \cite{reviews}. 
Heavy here means not Standard Model (SM) neutrinoÕs or $u$ and $d$ quarks, in some cases $s$ quarks, but mainly $c$ and $b$ quarks, as the $t$ is too heavy, decaying before it forms a hadron.

There are several ``problems" that the SM cannot explain. Baryogenesis,  the current dominance of matter over anti-matter requires, according to current models, much more CP violation than can be obtained from SM quark mixing \cite{baryo}. The SM also cannot explain dark matter affecting stars orbiting in galaxies \cite{dark},  and then there is the ``hierarchy problem," which can be summarized as our lack of understanding how to get from the Planck scale of Energy $\sim10^{19}$ GeV to the Electroweak Scale $\sim$100 GeV without fine tuning quantum corrections \cite{hier}.

\begin{figure}[ht]
\centering
\includegraphics[width=80mm]{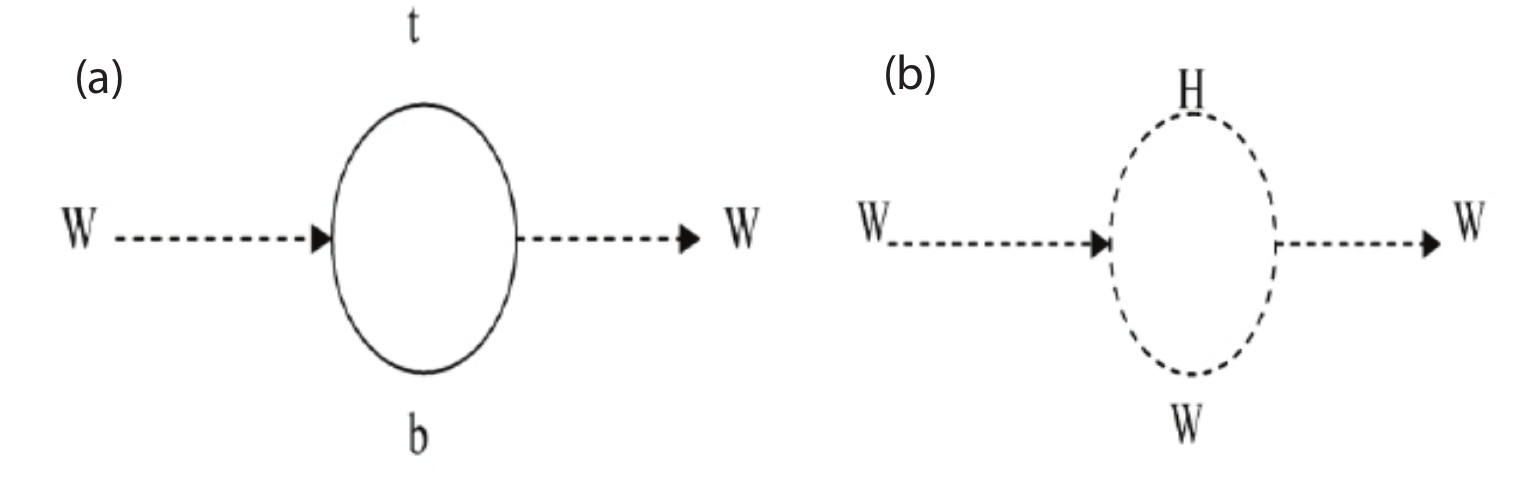}
\vspace{-2mm}\caption{Corrections to the $W$ mass due to (a) the $t$ quark and (b) the Higgs.} \label{quan-loop}
\end{figure}

In this paper I emphasize that the main purpose of HFP is to find and/or define the properties of physics beyond the SM. This is effective because HFP can probe large mass scales via virtual quantum loops. An example of the importance of such loops is extracting the Higgs mass. $W$ mass changes due the presence of $t$ are shown in Fig.~\ref{quan-loop}(a). The changes due to the Higgs are shown in Fig.~\ref{quan-loop}(b). 

\section{Limits on New Physics}

\begin{figure}[ht]
\centering
\includegraphics[width=80mm]{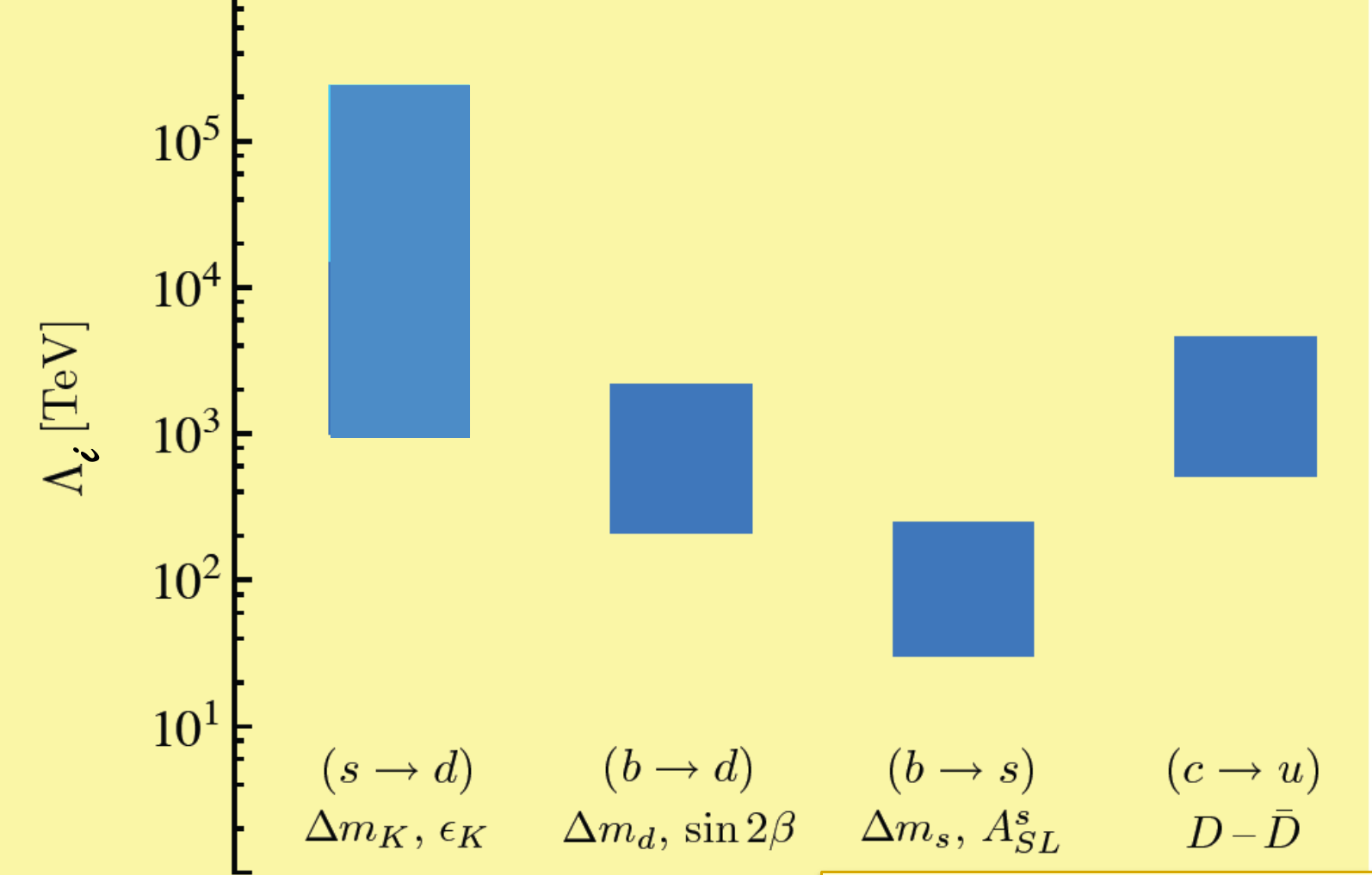}
\caption{New Physics reach in terms of mass scale ($\Lambda_i$) for various processes from \cite{INP}.} \label{NP-reach}
\end{figure}

Searches via quantum loops already have reached quite large mass scales given certain assumptions. Consider an effective Lagrangian given as
${\cal{L}}_{\rm eff} = {\cal{L}}_{\rm SM}+\frac{c_i}{\Lambda_i}{\cal{O}}_i,$ 
where ${\cal{O}}_i$ corresponds to new physics (NP) operators, and $\Lambda_i$ the mass scale. The coupling constants $c_i$ are taken as 1. Then the approximate range of different processes rules out NP in the regions shown in Fig.~\ref{NP-reach} \cite{INP}.

There are several ways to avoid these limits, however. (i) The new particles could have very large masses. (ii) The new particles could be degenerate in mass. (iii) The mixing angles in the NP sector are the same as in the SM; this is called Minimal Flavor Violation \cite{MFV}. Collectively these consideration already provide strong constraints on NP.

\begin{figure}[ht]
\centering
\includegraphics[width=135mm]{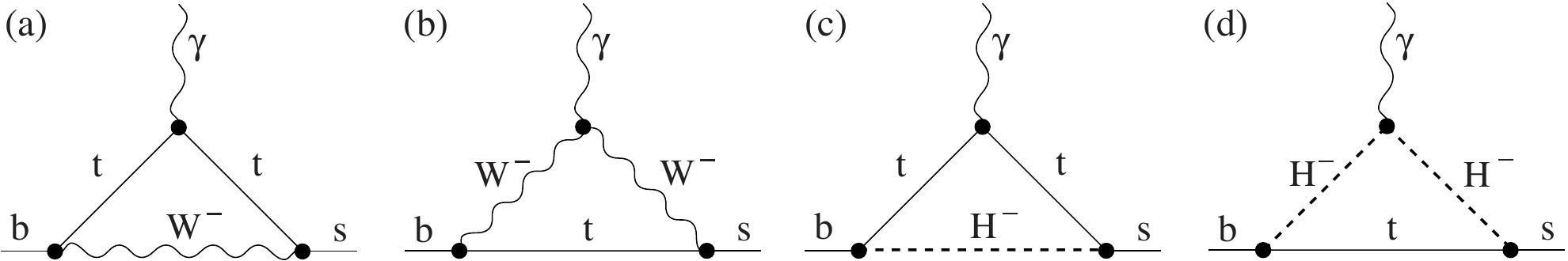}
\caption{ Decay diagrams for $b\to s\gamma$ (a) and (b) are SM while (c) and (d) show a possible NP process.} \label{sm}.
\end{figure}

\begin{figure}[!hbt]
\centering
\vspace{-5mm}\includegraphics[width=80mm]{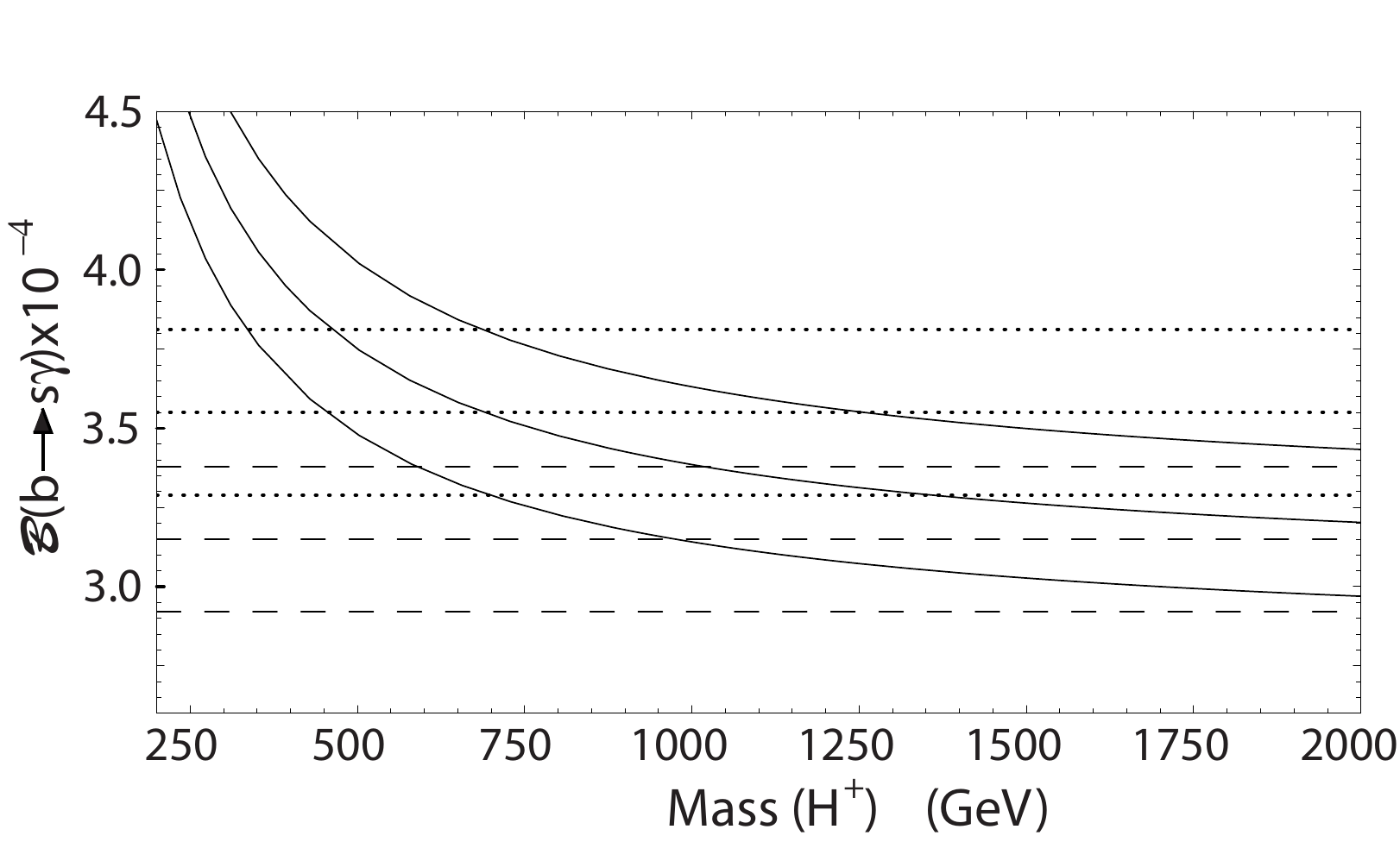}
\caption{The measured ${\cal{B}}(b\to s\gamma)$ shown as the central dotted line with upper and lower lines showing the uncertainty. The SM theory prediction is shown in the same manner using the dashed lines. The solid curves show the 2HDM with $\tan\beta$=2, from \cite{Misiak}.  } \label{MHc}.
\end{figure}

A specific example of constraints on NP is provided by the process $b\to s\gamma$. The SM diagrams for this process are shown in Fig.~\ref{sm}(a) and (b), while possible NP decays are shown in (c) and (d). The data are consistent with the SM prediction as shown in Fig.~\ref{MHc}  and rule out particular models of NP.  The central solid curve shown on the figure shows the prediction for the two-Higgs doublet model (2HDM) with $\tan\beta$=2. The outer curves show the model dependent error band. Comparison of the lower curve with the measurement restricts the charged Higgs mass to be greater than 316 MeV at 90\% confidence level (cl).

\begin{figure}[!htb]
\centering
\includegraphics[width=135mm]{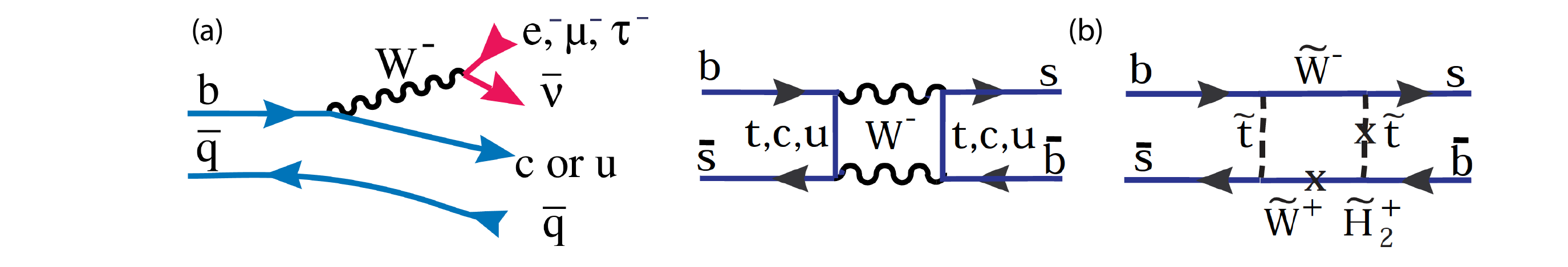}
\caption{ Decay diagram examples for (a) tree and (b) loop processes.} \label{tree-loop}
\end{figure}

It is often said that we have not seen NP, yet what we observe is the sum of SM plus NP. To set limits on NP we need to adopt a specific scenario. One way is to assume that ``tree level" diagrams are dominated by the SM while loop diagrams contain both SM and possibly NP.
Fig.~\ref{tree-loop} gives examples of these processes.

The CKM quark mixing matrix in the Wolfenstein formalism has four independent parameters, $A$, $\lambda$, $\rho$ and $\eta$ \cite{Wolf}. The values of $A$ and $\lambda$ have been measured and are about 0.8 and 0.225, respectively \cite{PDG}.  Most measurements involve an algebraic combination of $\rho$ and $\eta$ so the results are typically shown as bands on a $\eta-\rho$ plot. 

\begin{figure}[ht]
\centering
\hspace*{-8mm}\includegraphics[width=60mm]{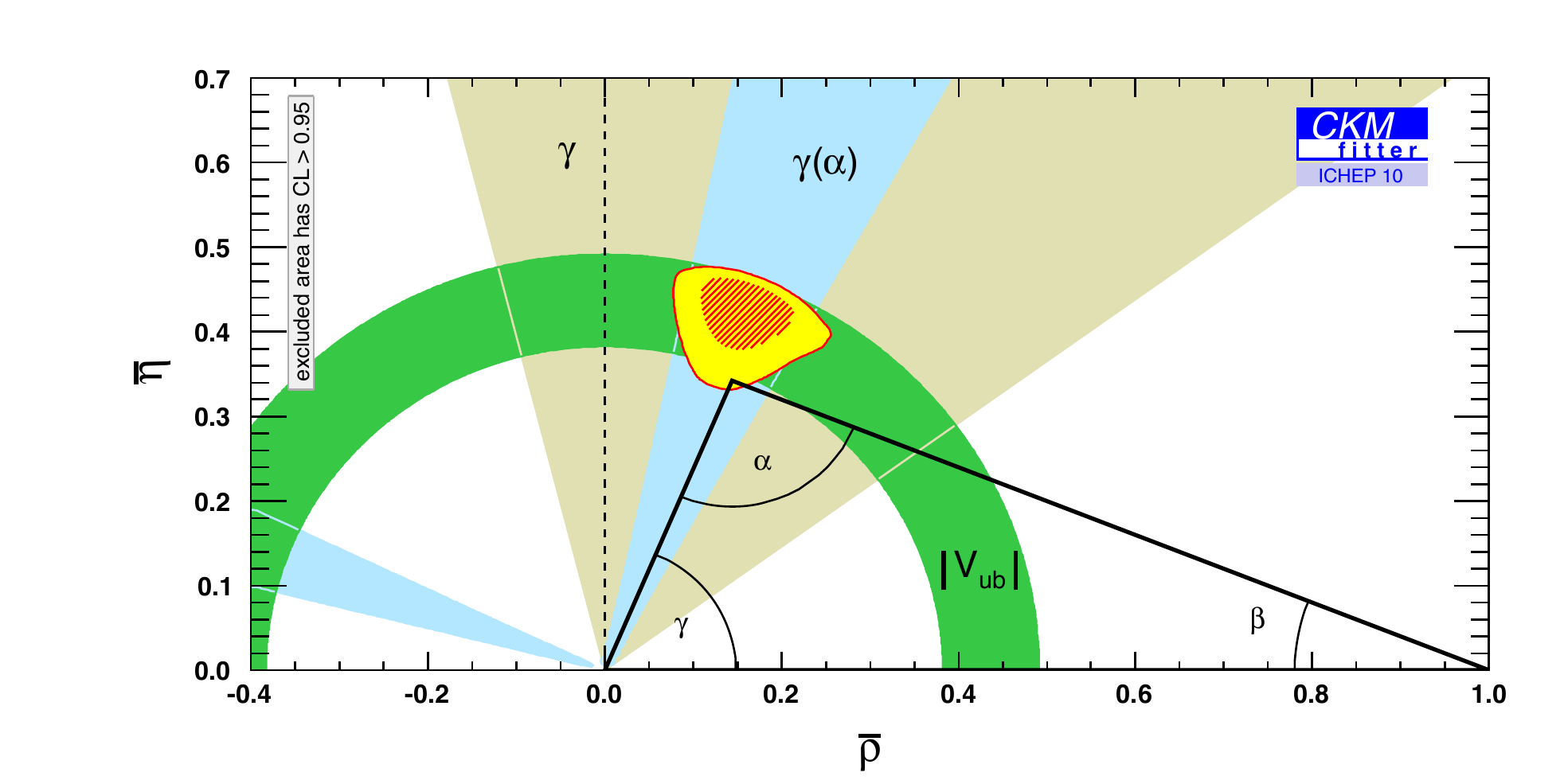}\includegraphics[width=60mm]{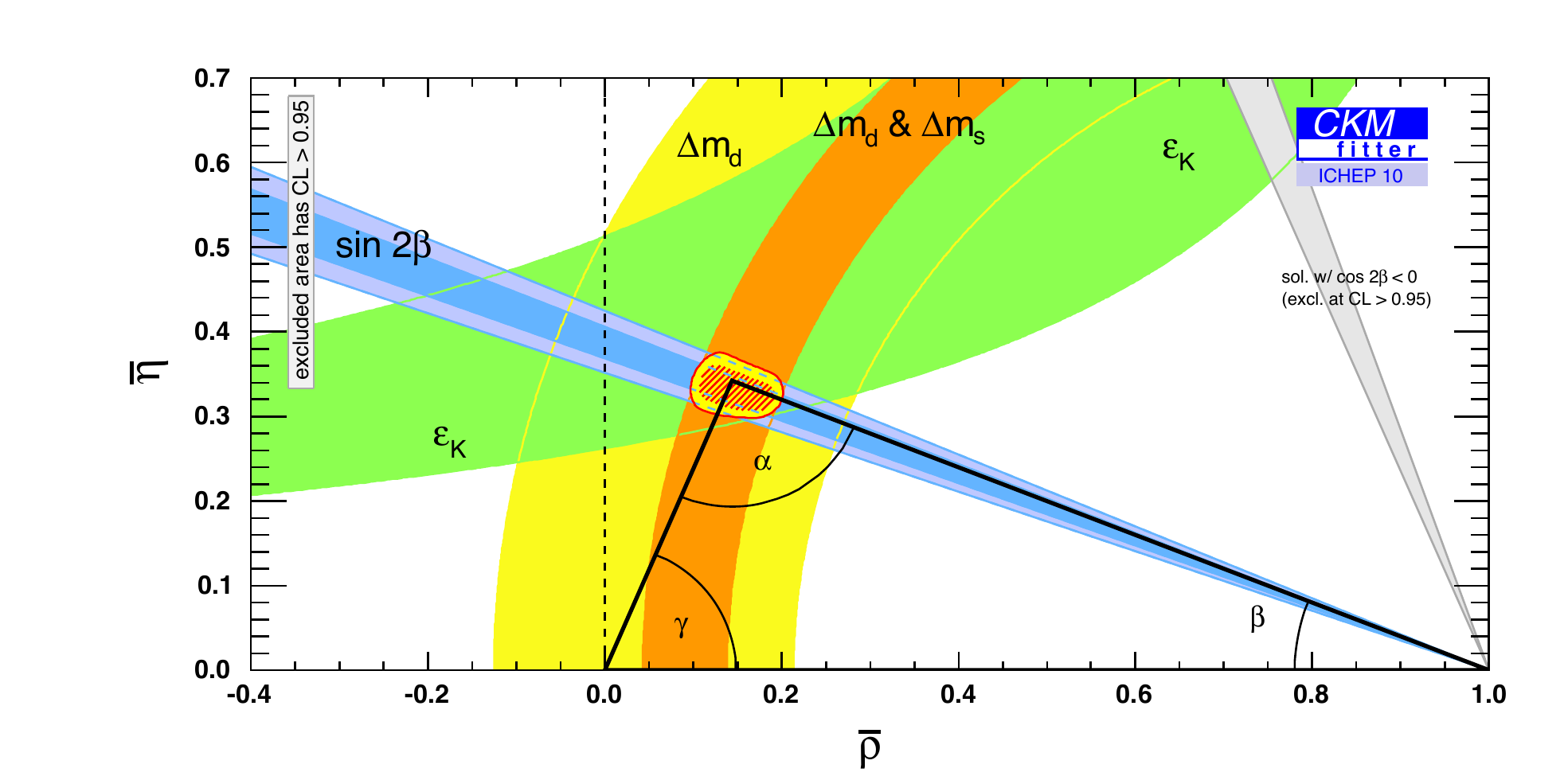}\includegraphics[width=60mm]{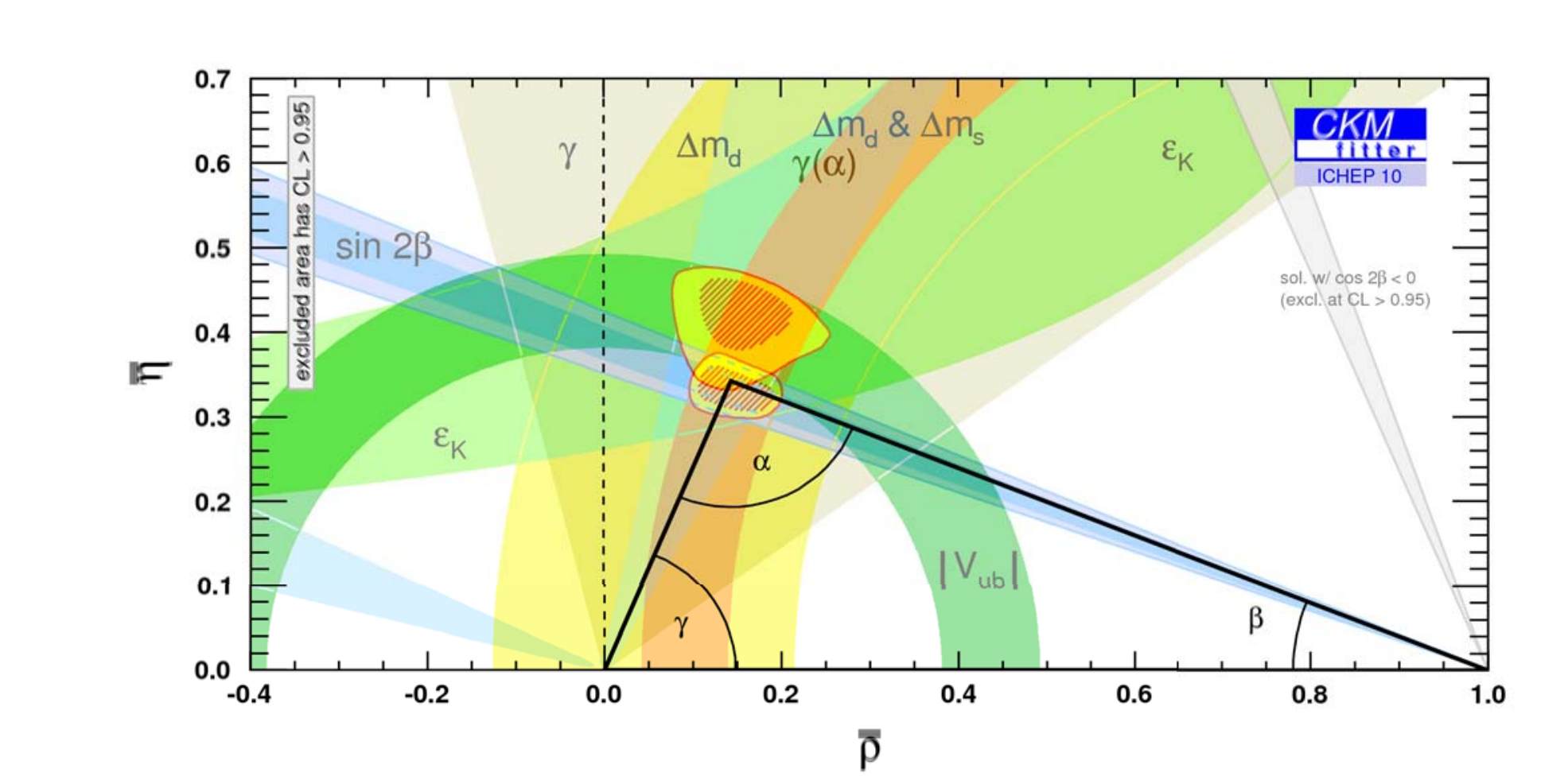}
\caption{ Constraints in the $\rho-\eta$ plane from (a) tree diagrams, (b) loop diagrams and (c) the overlay.} \label{rho-eta}
\end{figure}

Analysis of data involving only tree diagrams (Fig.~\ref{rho-eta}(a)) and loop diagrams (Fig.~\ref{rho-eta}(b)) is provided by the CKM fitter group \cite{CKMfitter}. An overlay of measurements shown in Fig.~\ref{rho-eta}(c) shows that the agreement is only at the 5\% confidence level leaving a lot of room for NP. 

%\section{Specific Topics of Interest}
%\subsection{\boldmath One Clear Problem: ${\cal{B}}(B^-\to\tau^-\overline{\nu})$ compared with $\sin2\beta$}

Closer investigation of the CKM fits reveals that the largest discrepancy occurs between the measured and predicted fit values of the branching fraction for $B^-\to\tau^-\overline{\nu}$  and the quintessential loop level process for CP violation in the $B^0$ system, resulting in $\sin2\beta$.
The process $B^-\to\tau^-\overline{\nu}$ proceeds via a tree level diagram in the SM where the $b$ annihilates with the $\overline{u}$ quark via a virtual $W^-$ that materializes as a $\tau^-\overline{\nu}$ pair. (This is the same diagram with different initial state quarks that describes $D_{(s)}^+\to \mu^+\nu$ decays.) The branching ratio is given by
\begin{equation}
{\cal{B}}(B^-\to\tau^-\overline{\nu})=\frac{G_F^2m_B\tau_B}{8\pi}m_{\tau}^2\left(1-\frac{m_{\tau}^2}
{m_{B}^2}\right)^2f^2_B\left|V_{ub}\right|^2,
\end{equation}
where $f_B$ is the decay constant determined theoretically, $|V_{ub}|$ is an important CKM element discussed below,
and the masses, lifetimes etc.. are very well measured.

\begin{figure}[hbt]
\centering
\includegraphics[width=60mm]{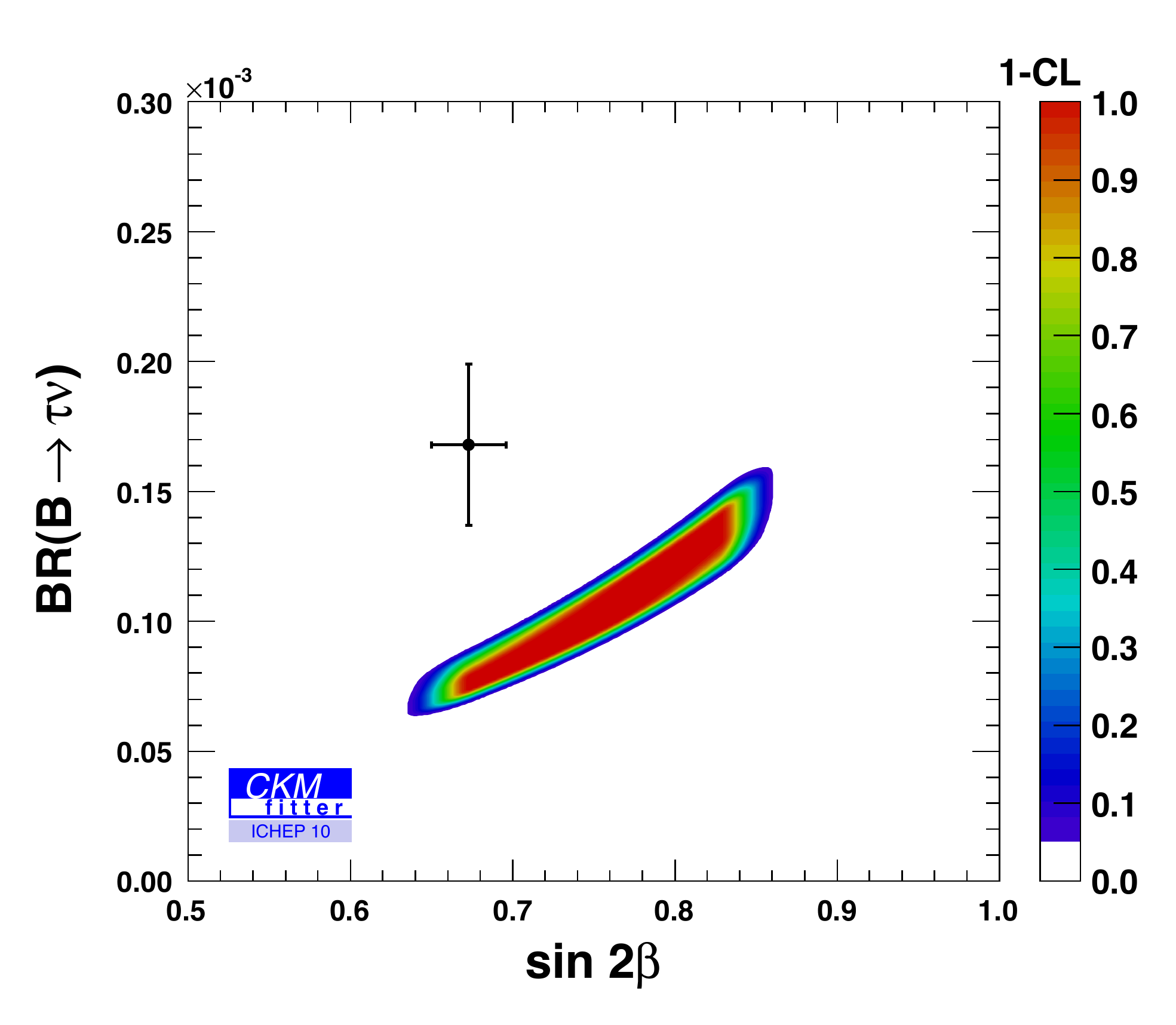}
\vspace{-4mm}
\caption{Measured versus predicted values of ${\cal{B}}(B^-\to\tau^-\overline{\nu})$ versus $\sin2\beta$ from the CKM fitter group.} \label{sin2b_Btaunu}
\end{figure}

 The point in Fig.~\ref{sin2b_Btaunu} shows the directly measured values, while the predictions from the fit without the direct measurements are also shown.  There is more than a factor of two discrepancy between the measured value of  ${\cal{B}}(B^-\to\tau^-\overline{\nu})$ and the fit value, somewhat more than a two standard deviation difference.
In a separate analysis Lunghi and Soni using different theoretical inputs claim a larger discrepancy \cite{Soni}. They draw a drastic conclusion reflected in the papers title: ``Demise of CKM and its aftermath."  A fourth generation based explanation is advocated.

\section{New Results}
There is a lot of new data from CDF, D0, BaBar, CLEOc, BES, BELLE, ATLAS, CMS and LHCb. It is not possible to cover all these excellent results even though Hassan Jawahery will cover the CP measurements.  Apologies to all of those whose results I do not cover. 
NP must affect every process; the amount in each process helps classify the NP  (``DNA footprint"). Some processes where the predictions in the SM have small errors and thus could reveal NP even with currently available data. I start with a discussion of $|V_{ub}|$.

\subsection{\boldmath Determination of $|V_{ub}|$}

Semileptonic $B$ decays are used to determine the value of $|V_{ub}|$. The decay process is shown in Fig.~\ref{tree-loop}(left). Both inclusive decays, where only the final state lepton is detected, and exclusive decays, primarily the $\pi\ell^-\overline{\nu}$ final state are used.  These results, however, do not agree. 

\begin{figure}[htb]
\centering
\includegraphics[width=135mm]{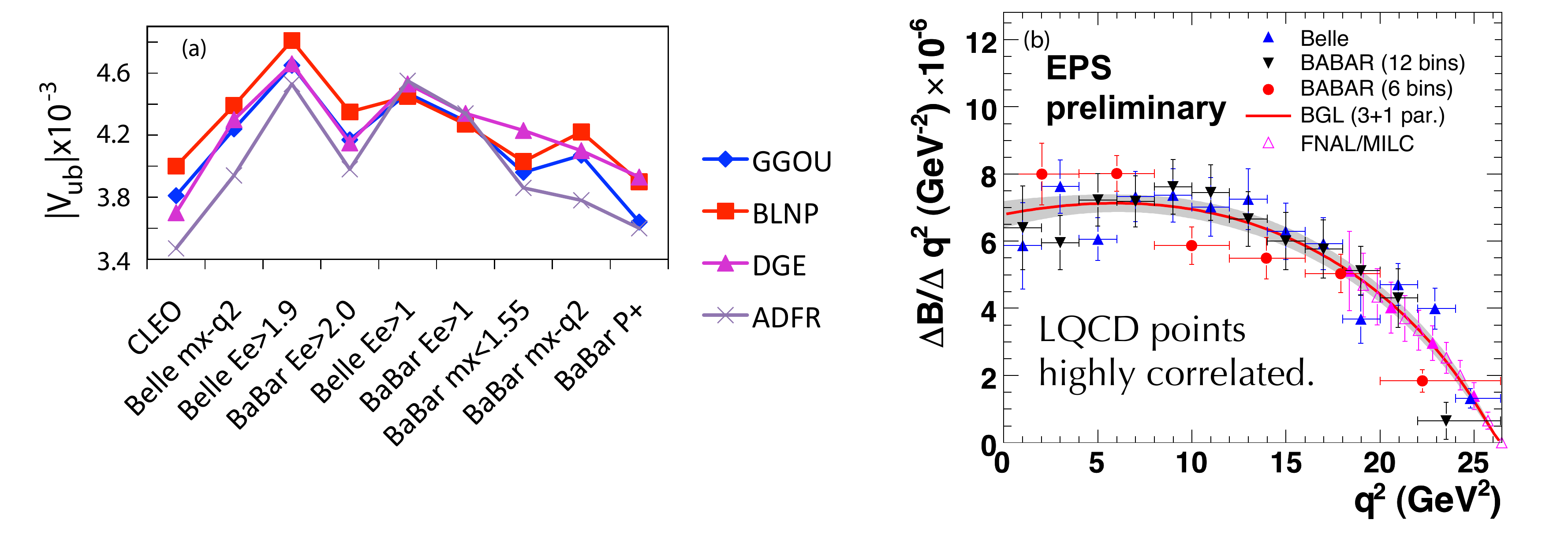}
\vspace{-0.2cm}
\caption{(a) Values of $|V_{ub}|$ determined from different inclusive measurements \cite{Kowa}. (b) The $q^2$ dependence for $\overline{B}^0\to\pi^+\ell^-\overline{\nu}$ from several different experiments and from
a Lattice QCD model \cite{Ur}.} \label{Vub}
\end{figure}

\begin{figure}[htb]
\centering
\vspace{-0.5cm}\hspace*{-2.5cm}\includegraphics[width=230mm]{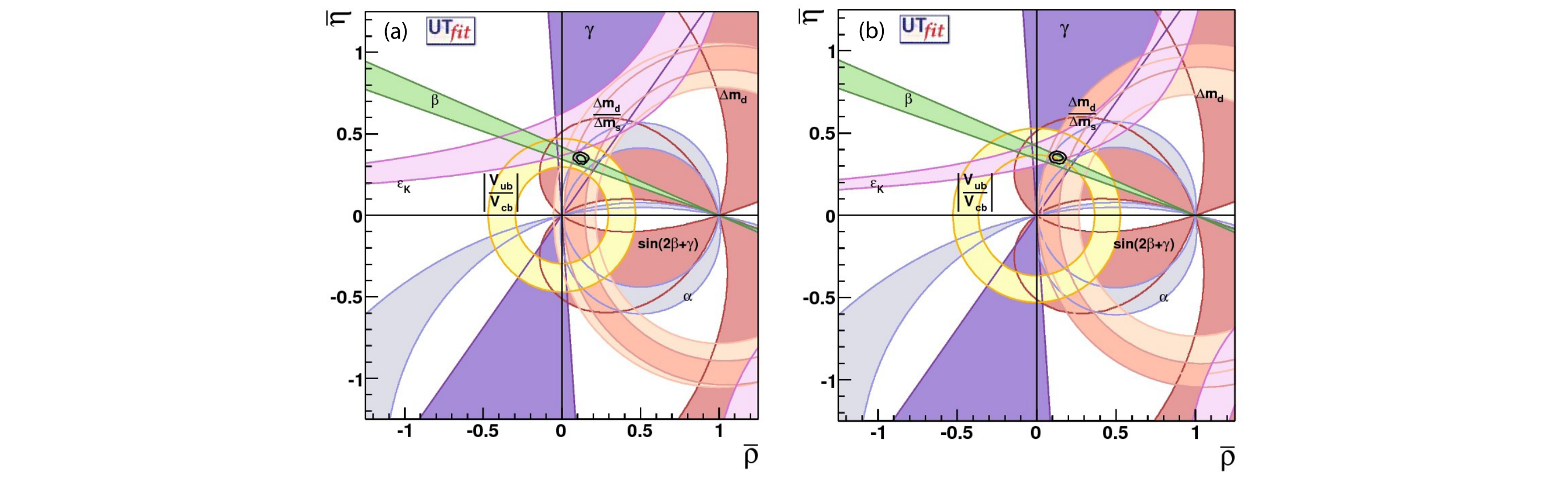}
\vspace{-0.6cm}
\caption{(a) UT fit results using the exclusive determination of $|V_{ub}|$, and (b) inclusive \cite{UT}.} \label{Vub-ex-vs-in}.
\end{figure}

Inclusive values are determined in the context of different models using different kinematic cuts to enhance various phase space regions, and applying the heavy quark expansion to extract the relevant hadronic matrix element \cite{HQE}. Kowalewski quotes an average value $|V_{ub}|=4.25\pm 0.15\pm 0.20$ \cite{Kowa}, based on the results shown in Fig.~\ref{Vub}(a). The exclusive measurements are mostly based on the $\overline{B}^0\to\pi^+\ell^-\overline{\nu}$ decay mode. Theoretical models predict the branching fraction in specific ranges of four-momentum transfer, $q^2$, between the $B$ and the $\pi$. The data and some theoretical points from Lattice QCD are shown in Fig.~\ref{Vub}. Urquijo quotes an average value $|V_{ub}|=3.25\pm 0.12\pm 0.28$ \cite{Ur}, leading to a  $\sim$25\% discrepancy with the exclusive results. 
This irritating and somewhat long term problem doesn't have a major effect on the ${\cal{B}}(B^-\to\tau^-\overline{\nu}) - \sin2\beta$ discrepancy, nor on the overall difference between tree and loop diagrams as can be seen by the separate fits done to the overall CKM measurements by the UT fit group shown in Fig.~\ref{Vub-ex-vs-in} \cite{UT}.

\begin{figure}[ht]
\centering
\hspace*{-1.5cm}\includegraphics[width=90mm]{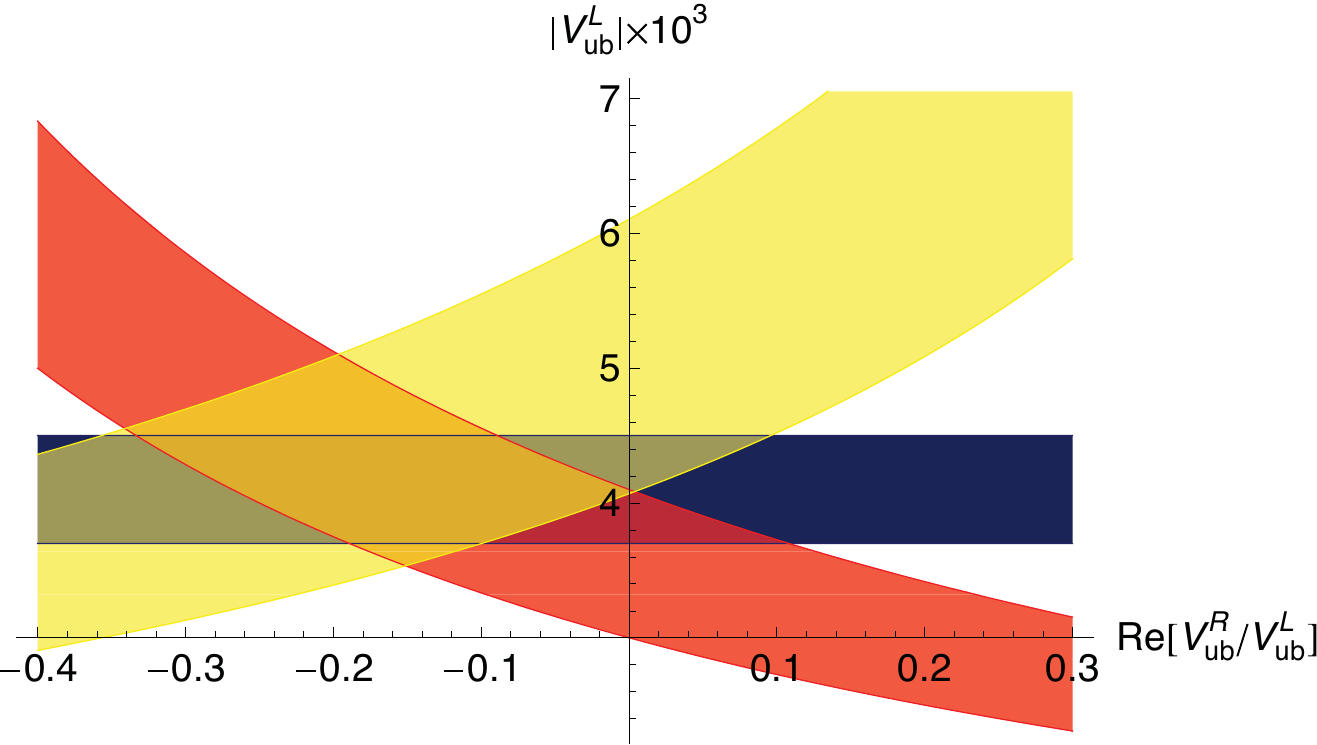}
%\vspace{-8.2cm}
\caption{The correlation between $\left|V_{ub}^L\right|$ and $Re\left[V_{ub}^R/V_{ub}^L\right]$ \cite{Criv}.} \label{Crivellin}
\end{figure}

While the differences alluded to here can have several sources including theoretical or experimental errors, it is also possible that NP is the cause. A neat explanation is to allow a right-handed current coupling. Then the decay rate for $B\to\pi\ell\nu$ is proportional to $\left|V_{ub}^L+V_{ub}^R\right|^2$, the rate for $B\to\tau\nu$ $\propto\left|V_{ub}^L-V_{ub}^R\right|^2$, and the rate for $B\to X_u\ell\nu$ (exclusives)  $\propto\left|V_{ub}^L\right|^2+\left|V_{ub}^R\right|^2$. A plot of the correlation between $\left|V_{ub}^L\right|$  and $Re\left[V_{ub}^R/V_{ub}^L\right]$ done by Crivellin is shown in Fig.~\ref{Crivellin} \cite{Criv}. Allowing a 15\% right-handed current brings all three measurements into agreement. This of course would be NP at the tree level, and should also show up in other modes  \cite{Buras-Vub}.

\subsection{Forward-Backward Asymmetry in $\overline{B}^0\to \overline{K}^{*0}\mu^+\mu^-$ Decays}
\begin{figure}[ht]
\centering
\vspace{-6mm}\includegraphics[width=90mm]{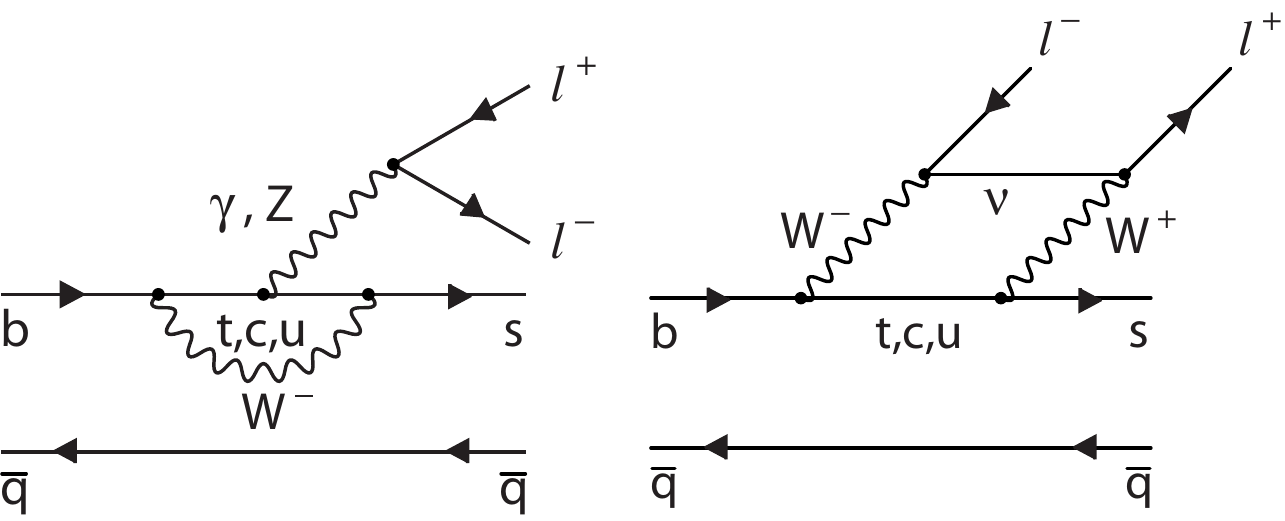}
%\vspace{-4.2cm}
\caption{$A_{FB}$ in  $\overline{B}^0\to \overline{K}^{*0}\mu^+\mu^-$ decays.} \label{kstarmumu-diag}
\end{figure}

\begin{figure}[ht]
\centering
\vspace{-1.4cm}\includegraphics[width=100mm]{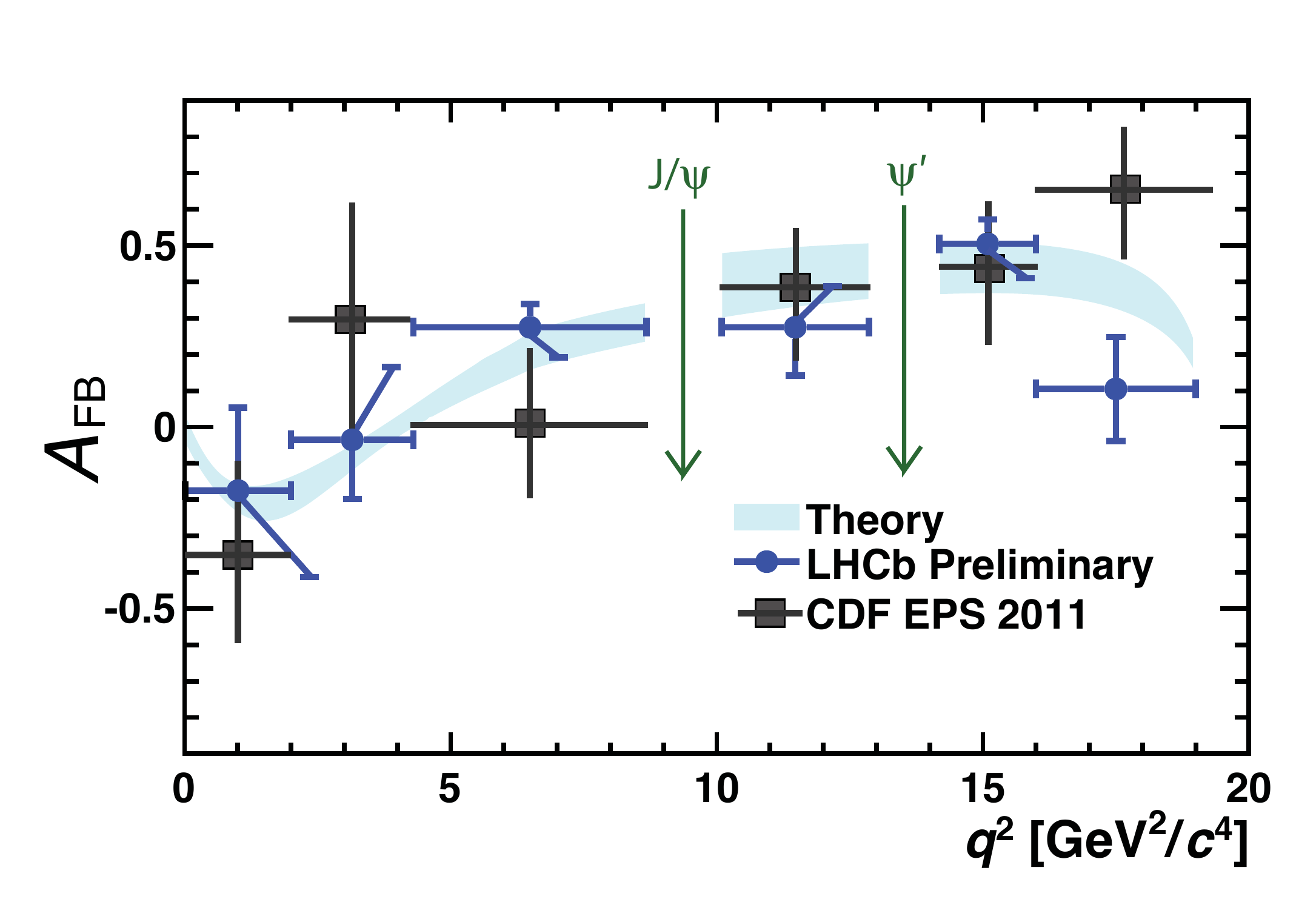}
\vspace{-.2cm}
\caption{CDF and LHCb measurements of $A_{FB}$ in  $\overline{B}^0\to \overline{K}^{*0}\mu^+\mu^-$ decays compared with the SM prediction \cite{BHD}.} \label{Kstarmumu-both}
\end{figure}

Next we turn to a process that could explicitly show the effects of NP. The decay $\overline{B}^0\to \overline{K}^{*0}\mu^+\mu^-$ is similar to $B\to K^*\gamma$, but there are more SM diagrams involving loops as shown in Fig.~\ref{kstarmumu-diag}.  There are several variables that can be examined. One that can be accurately predicted in the SM is the forward-backward asymmetry of the dimuon pair, $A_{FB}$, as a function of $q^2$.  Measurements of BaBar, Belle and CDF had indicated, with low statistics, that $A_{FB}$ was larger than the SM prediction especially at $q^2$ close to zero \cite{Ksmm-old}. New results from CDF using 6.8 fb$^{-1}$ \cite{Ksmm-CDF} and LHCb using 0.3 fb$^{-1}$\cite{Ksmm-LHCb} now are quite consistent with the SM prediction as shown in Fig.~\ref{Kstarmumu-both}. 

\subsection{Determination of the $b$-hadron Production Fractions at the LHC}
While absolute branching fractions of $B^-$ and $\overline{B}^0$ mesons have been determined accurately by experiments operating at the $\Upsilon(4S)$ resonance \cite{PDG}, an important part of the heavy flavor physics program involves measuring absolute branching ratios of $\Bs$ mesons. To determine this ratio LHCb uses two techniques, one using fully reconstructed hadronic decays, and the other partially reconstructed semileptonic decays. In the hadronic modes two ratios are formed, $\Gamma(\overline{B}^0\to D^+K^-)/\Gamma(\Bs\to D_s^+\pi^-)$ and $\Gamma(\overline{B}^0\to D^+\pi^-)/\Gamma(\Bs\to D_s^+\pi^-)$. Each one then is used with a theoretical correction based on factorization \cite{Fleischer} to extract the ratio of $\Bs$ to $\overline{B}^0$ production, $f_s/f_d=0.253\pm 0.017\pm 0.017\pm 0.020$, where the uncertainties are respectively statistical, systematic and theoretical \cite{had-frac}. 

In the semileptonic method the individual charm hadrons $D^0$, $D^+$, $D_s^+$ and $\Lambda_c^+$ are found, required to form a detached vertex with a $\mu^-$, and not to have their momentum vector point at the primary vertex which is characteristic of $b$ hadron decay.  Most of the decays to $D^0$ and $D^+$ result from either $\overline{B}^0$ or $B^-$ decays,  but there is a component of $\Bs$ decays into these mesons plus a kaon, and $\Lambda_b^0$ decays into a $D^0$ or a $D^+$ plus a proton or neutron. These cross-feeds are measured. Similarly, most of the $D_s^+$ result from $\Bs$ decays with cross-feeds from the other modes. The resulting ratio measured ratio is actually $f_s/(f_u+f_d)$. Isospin conservation is used to set $f_d=f_u$. Then $f_s/f_d=0.268\pm 0.008^{+0.022}_{-0.020}$ \cite{semi-fracs}. The theoretical uncertainties are very small. In fact, the dominant systematic uncertainty of 5.5\% is due to the errors on the charm hadron branching fractions. The combination of the two methods to determine the fractions is dominated by the semileptonic result, and LHCb quotes $f_s/f_d=0.267^{+0.021}_{-0.020}$ \cite{combo-fracs}. The 8\% precision achieved here is sufficient to allow a relative precise determination of $\Bs$ branching fractions.

\begin{figure}[hbt]
%\vspace{-.8cm}
\center
\includegraphics[width=6.7in]{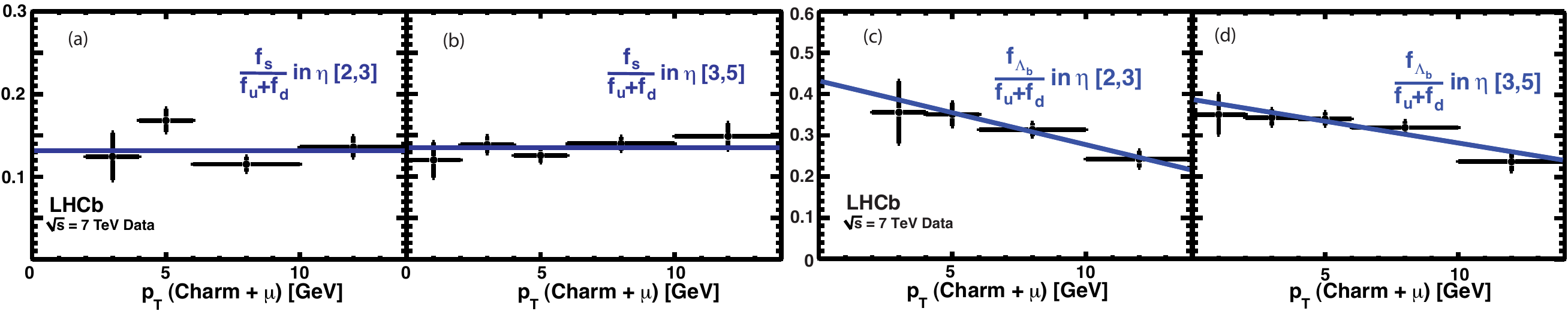}
%\vspace{-1.1cm}
\caption{\label{bfracs2} The $b$-hadron fraction $f_s/(f_u+f_d)$ in (a) and (b) as a function of $p_T$ in two $\eta$ intervals, and 
similar distributions for $f_{\Lambda_b}/(f_u+f_d)$  in (c) and (d), both from 3 pb$^{-1}$ of data. }
\end{figure}

The semileptonic method also provides a result for the $\Lambda_b^0$ production fraction. It is large, $\approx$30\%. Both $f_s/(f_u+f_d)$ and $f_{\Lambda_b}/(f_u+f_d)$ are constant with $\eta$, where $\eta=-\log\tan\theta/2$, and $\theta$ is the $b$ production angle with respect to the beam direction.  The variable $p_T$ here is defined as the vector sum of the charmed hadron and muon momenta.  It is found that $f_{\Lambda_b}/(f_u+f_d)$ has a large variation with $p_T$ at LHCb, while $f_s/(f_u+f_d)$ is constant \cite{semi-fracs}. Both results are shown in Fig.~\ref{bfracs2}.

\subsection{\boldmath Measuring the Branching Fraction of $\Bs\to\mu^+\mu^-$}
\begin{figure}[ht]
\centering
\includegraphics[width=125mm]{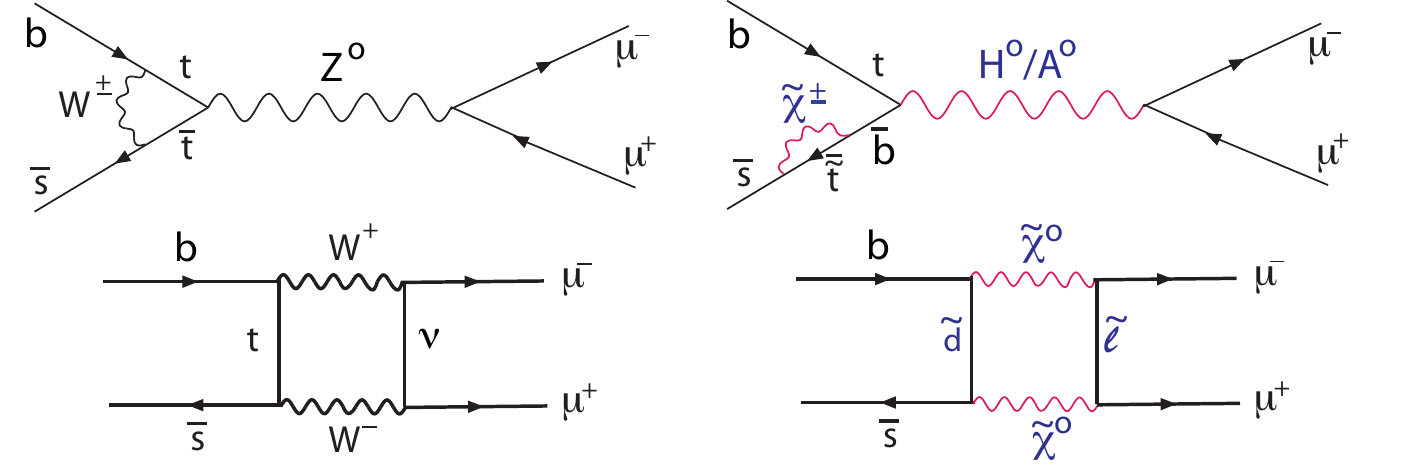}
%\vspace{-4.2cm}
\caption{Feynman diagrams for $\Bs\to\mu^+\mu^-$ from (left) the SM and (right) supersymmetry (MSSM).} \label{Bsmumu}
\end{figure}

$\Bs\to\mu^+\mu^-$ is highly suppressed in the SM with a predicted branching fraction of $(3.2\pm0.2)\times 10^{-9}$ \cite{Buras-mm}, and it is possible for NP to increase the rate by large factors \cite{Thybmumu}. The SM diagram and one possible NP diagram from supersymmetry are shown in Fig.~\ref{Bsmumu}.

\begin{figure}[htb]
\centering
\includegraphics[width=135mm]{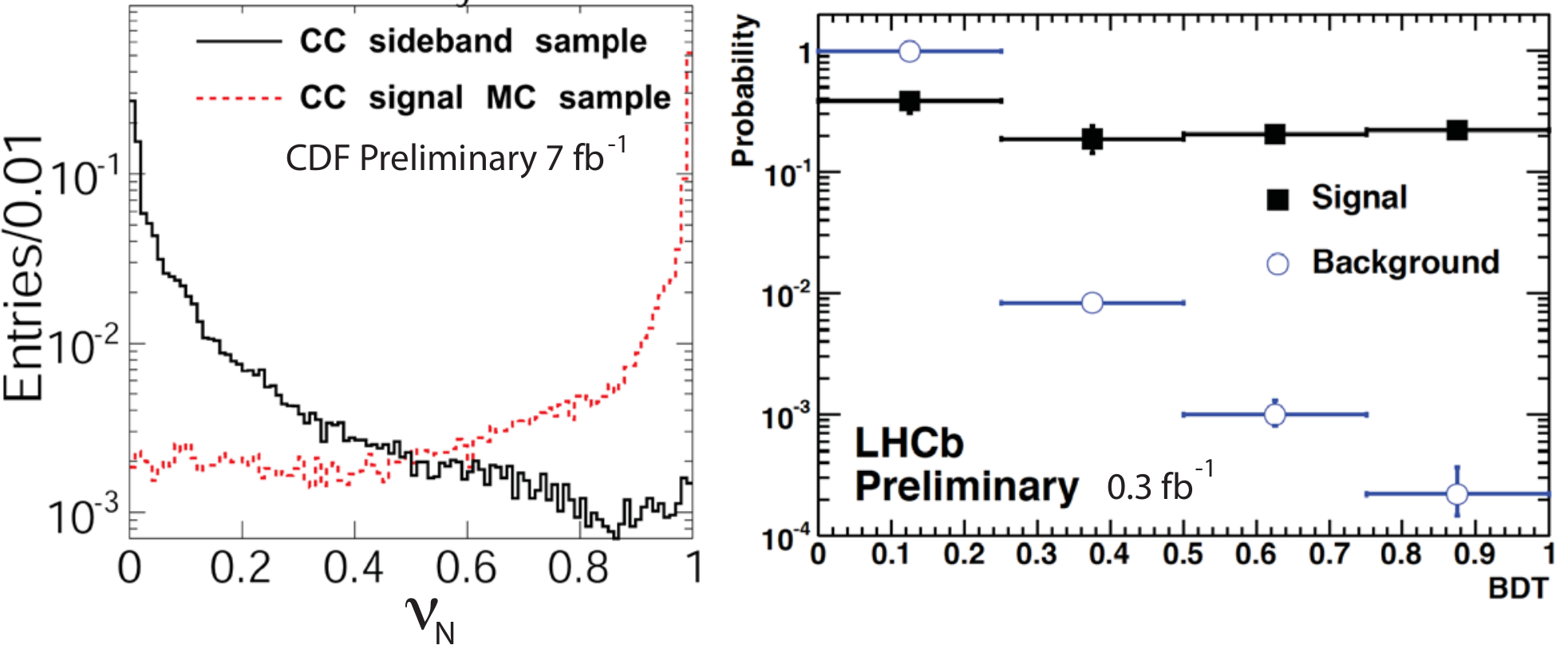}
\vspace{-.2cm}
\caption{Multivariate analysis variable distributions from simulated signals and data based backgrounds from (left) CDF, and (right) LHCb.} \label{mumu-disc}
\end{figure}

Techniques used to enhance signal over background in the search for this rare decay can be optimized on data because there are topologically identical and much more prolific processes $B\to h^+h^-$, where $h$ is either a pion or a kaon.  The variables used to discriminate include $B$ impact parameter, $B$ lifetime, $B$ transverse momentum, $B$ isolation, track isolation, minimum impact parameter of tracks, etc..  Subsequently muon identification is added. The multivariate analyses of CDF and LHCb produce the simulated distributions of signal, and background determined from data, in terms of the internally defined variables shown in Fig.~\ref{mumu-disc}.  Signal efficiency is high at large values of the CDF variable $\nu_N$, where the backgrounds are minimal. For LHCb, the BDT variable is nearly flat in signal efficiency (by design), while the background falls off quickly above BDT of 0.5. 

\begin{figure}[ht]
\centering
\includegraphics[width=125mm]{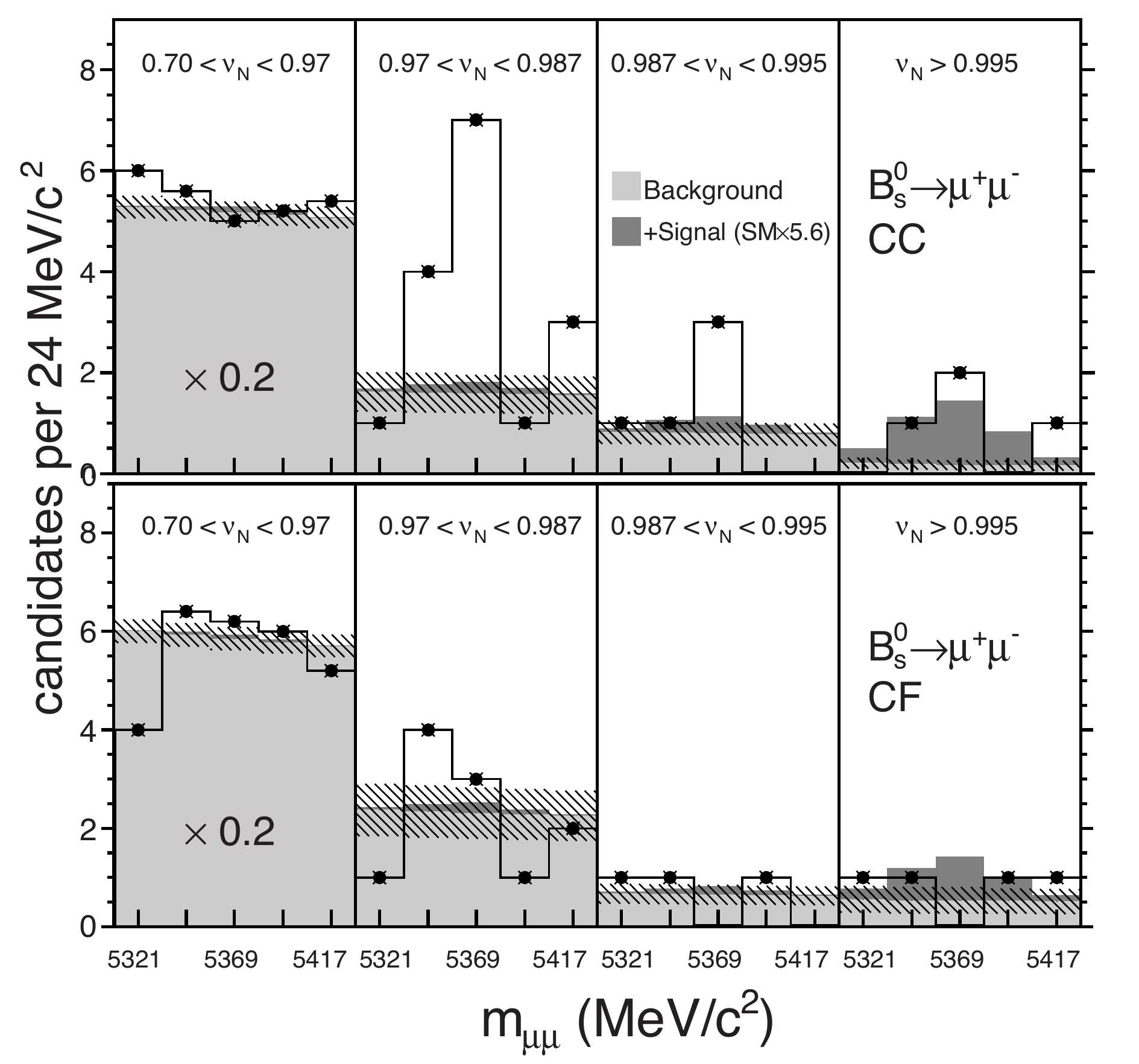}
\vspace{-.4cm}
\caption{The observed number of events (open histogram with points)  
    compared to the total expected background (light grey) and its uncertainty (hatched) 
    in bins of dimuon invariant mass.  The top and bottom rows show the results in the 
    $\Bs$ mass signal region for the CC and CF channels, respectively.  
    The results for the first 5 $\nu_N$ bins are combined (and scaled by $0.2$) while the results
    for the last three bins are each shown separately. Also shown is the expected contribution
    from signal events (dark gray) using the fitted branching fraction, which is 5.6 times 
    the expected SM value.} \label{Bmm-CDF}
\end{figure}

 CDF recently reported a tantalizing observation when searching for a signal in $\Bs\to\mu^+\mu^-$ using 7 fb$^{-1}$ \cite{Bmumu-CDF}.  The CDF data are separated in two samples, one where both muons are in the central region of the detector (CC) and another where one muon is in the central region and the other is in the forward region (CF). The CF sample has 72.5\% of the efficiency of the CC sample, and the backgrounds expected to be about a factor of 1.5 larger in the 3 highest $\nu_N$ bins \cite{Kuhr}. 
At the SM level 1.9 signal events are expected in the sum of both samples. 
The CDF result is shown in Fig.~\ref{Bmm-CDF}.
CDF claims an excess. It appears that it is mainly in the CC sample while none is visible in the CF sample. Averaging both samples CDF quotes a central value of ${\cal{B}}(\Bs\to\mu^+\mu^-)=1.8^{+1.1}_{-0.9}\times 10^{-8}$, 5.6 times the standard model value. The probability for background plus the SM rate to give this result is 2.9\%. Thus, CDF quotes a ``two-sided" limit $4.6\times 10^{-9}<
{\cal{B}}(\Bs\to\mu^+\mu^-)<3.9 \times 10^{-8}$. When a two-sided limit is quoted, I view the result as not being statistically significant.
Nevertheless, seeing if there is an excess rate in this channel is very important.
LHCb and CMS have provided such data. 

\begin{figure}[ht]
\centering
\includegraphics[width=160mm]{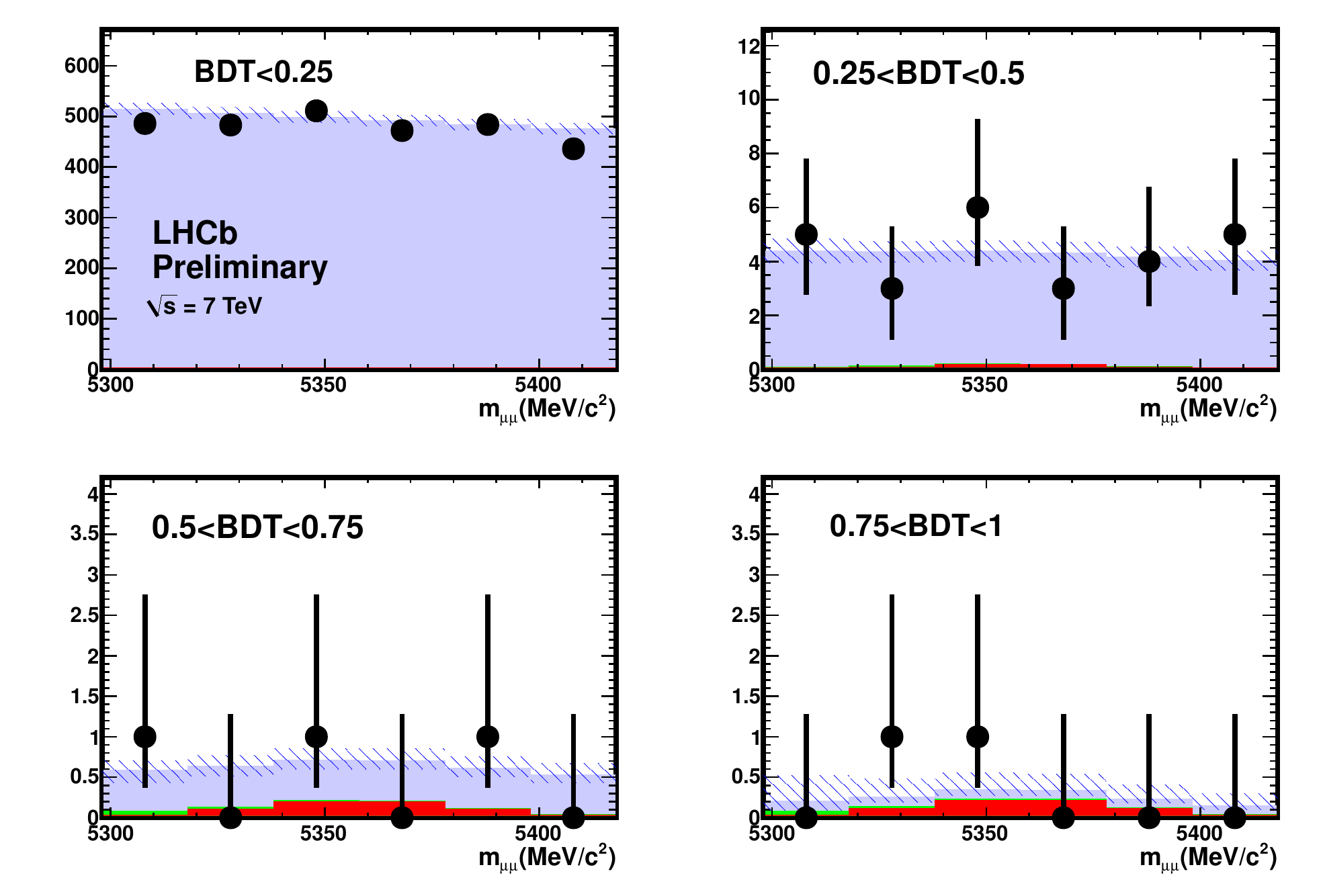}
\vspace{-0.2cm}
\caption{Distribution of selected dimuon events from LHCb as a function of invariant mass. The black dots are data, the light blue histogram shows the contribution of the combinatorial background with its uncertainty (hatched area). The
almost invisible green histogram shows the contribution of the $h^+h^-$ background, and the red 
histogram the contribution of signal events according to the SM rate.} \label{PlotBs}
\end{figure}

The LHCb data using 0.34 fb$^{-1}$ is shown in Fig.~\ref{PlotBs} and listed in Table~\ref{tab:LHCbmumu} \cite{LHCb-Bmm}.
LHCb does not observe an excess above the SM level. In the two BDT signal bins LHCb expects to see 5.1 events and sees 5.
CMS performs a cut based analysis using 1.14 fb$^{-1}$ \cite{CMS-Bmm}. They have two samples, one where both muons are in the barrel and the  other where one muon is in the endcaps or both are in the endcaps. Their results are presented in Table~\ref{tab:CMSbmumu}. CMS also does not observe an excess over SM expectations.

\begin{table}[ht]
\begin{center}
\caption{The LHCb results for $\Bs\to\mu^+\mu^-$ in bins of BDT. The {\bf bold} numbers show the expected versus observed
numbers in the two signal bins using the SM expectation for the signal.}
\begin{tabular}{|l|c|c|c|c|}
\hline & \textbf{BDT$<$1/4} & \textbf{1/4$<$BDT$<$1/2} & \textbf{1/2$<$BDT$<$3/4} &
\textbf{3/4$<$BDT$<$1}\\
\hline \# expected bkgrd& 2968$\pm$69 & 25.0$\pm$2.5 & 2.99$\pm$0.89& 0.66$\pm$0.40 \\
\hline \# expected signal& 1.26$\pm$0.13 & 0.61$\pm$0.6 & 0.67$\pm$0.0.07& 0.72$\pm$0.07 \\
\hline Sum expected & 2969$\pm$69 & 25.6$\pm$2.5 & {\bf 3.66$\pm$0.89}& {\bf 1.38$\pm$0.41} \\
\hline Observed & 2872 & 26 &{\bf 3}& {\bf 2} \\
\hline
\end{tabular}
\label{tab:LHCbmumu}
\end{center}
\end{table}

\begin{table}[hbt]
\begin{center}
\caption{The CMS results for $\Bs\to\mu^+\mu^-$.}
\begin{tabular}{|l|c|c|}
\hline & \textbf{Barrel} &\textbf{ Endcap}\\
\hline \# expected combinatoric bkgrd& 0.60$\pm$0.35 & 0.80$\pm$0.40\\
\hline \# expected $B\to h^+h^-$ bkgrd& 0.07$\pm$0.02 & 0.04$\pm$0.01\\
\hline \# expected signal& 0.80$\pm$0.16 & 0.36$\pm$0.07 \\
\thickhline Sum expected & 1.47$\pm$0.39 & 1.20$\pm$0.41 \\
\hline Observed & 2 & 1  \\
\hline
\end{tabular}
\label{tab:CMSbmumu}
\end{center}
\end{table}

\begin{figure}[ht]
\centering
\includegraphics[width=70mm]{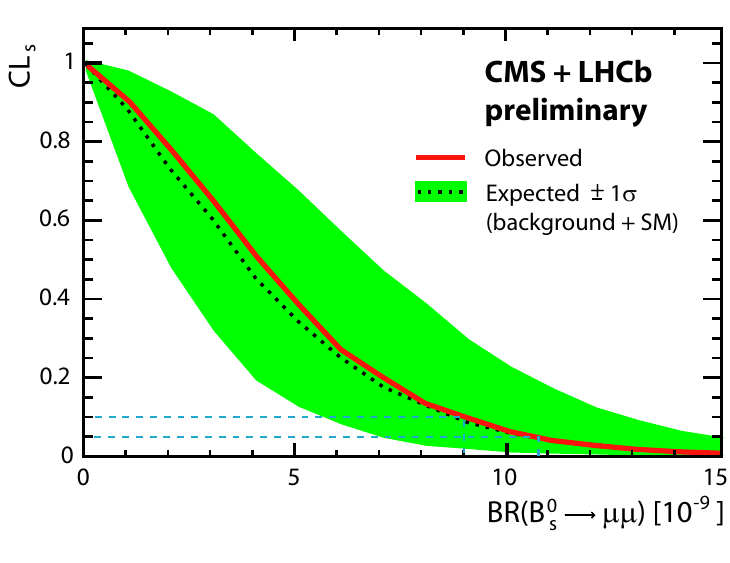}
\vspace{-0.4cm}
\caption{The observed (solid curve) and expected for background-only (dotted curve)
CLs values as a function of ${\cal {B}}(\Bs\to\mu^+\mu^-)$ . The green shaded area contains the $\pm 1\sigma$ interval
of possible results compatible with the expected value, when both background plus SM signal is observed.} \label{Bsmumuul}
\end{figure}

Table~\ref{tab:ul-combo} lists both the individual upper limits from LHCb and CMS, and the combined result \cite{combo}.
\begin{table}[hbt]
\begin{center}
\caption{Upper limits from LHCb, CMS and their combination for ${\cal{B}}(\Bs\to\mu^+\mu^-)$}
\begin{tabular}{|l|c|c|c|}
\hline                                    & LHCb $\times 10^{-8}$&  CMS  $\times 10^{-8}$& Combination $\times 10^{-8}$\\
\hline Upper limit @ 95\% cl &   1.5  &  1.9 & 1.1\\
\hline Upper limit @ 90\% cl &    1.2 &  1.6 &  0.9\\
\hline
\end{tabular}
\label{tab:ul-combo}
\end{center}
\end{table}
These limits are substantially smaller than the CDF central value, and corresponds to 2.8 times the SM prediction at 90\% cl. The confidence levels (CLs) versus ${\cal {B}}(\Bs\to\mu^+\mu^-)$ is shown in Fig.~\ref{Bsmumuul}. The combined data have a 0.3\% probability of being consistent with the central value of the CDF result. The combined LHC results present severe limits on NP models, yet there is still a lot of room for NP to appear before we reach the SM level \cite{Nonewphys}.

\afterpage{\clearpage}\newpage
\subsection{\boldmath First Observations of $\Bs\to J/\psi f_0(980)$ and $\Bs\to J/\psi f'_2(1525)$}
\begin{figure}[ht]
\centering
\includegraphics[width=165mm]{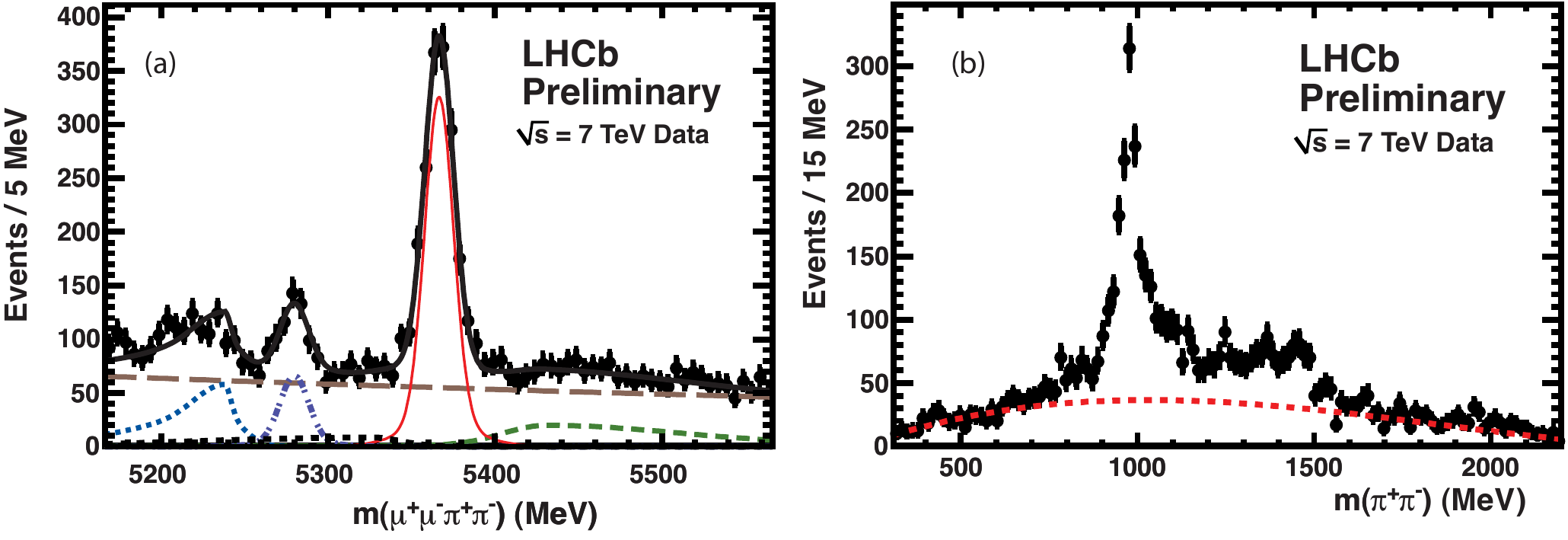}
%\vspace{-4.2cm}
\caption{(a) The invariant mass of $J/\psi\pi^+\pi^-$ combinations when the $\pi^+\pi^-$ pair is required to be within $\pm$90 MeV of the $f_0(980)$ mass. The data have been fit with a signal double-Gaussian function and several background functions. The thin (red) solid line shows the signal, the long-dashed (brown) line the combinatoric background, the dashed (green) line the $B^+$ background, the dotted (blue) line the $\overline{B}^0\to J/\psi \overline{K}^{*0}$ background, the dash-dot line (purple) the ${B}^0\to J/\psi \pi^+\pi^-$ background, the very small in level dotted line (black) the sum of $B_s\to J/\psi \eta'$ and $J/\psi\phi$ backgrounds, and the thick-solid (black) line the total. (b) The invariant mass of $\pi^+\pi^-$ combinations (points) and a fit to the  $\pi^+\pi^+ +\pi^-\pi^-$ data (dashed line) for events in the $\Bs$ signal region.} \label{Jpsif0}
\end{figure}

Time dependent measurements of CP violation in the $B_s^0-\Bs$ system have been based on $\Bs\to J/\psi\phi$ decays. It was suggested by Stone and Zhang that such determinations could be affected if there was a S-wave $K^+K^-$ component under the $\phi$ whose level could be $\sim$10\% \cite{Stone-Zhang}.
Furthermore it was predicted that the S-wave could manifest itself as $f_0(980)$ state and
\begin{equation}
R =\frac{{\cal{B}}(B_s^0\to J/\psi f_0,~f_0\to \pi^+\pi^-)}
{{\cal{B}}(B_s^0\to J/\psi \phi,~\phi\to K^+K^-)}=20\%.
\end{equation}
The $J/\psi f_0$ state is pure CP odd and could also be used to measure CP violation.
LHCb searched for and made the first observation of $J/\psi f_0(980)$ decays \cite{Jpsif01st} using a 36 pb$^{-1}$ data sample that was subsequently confirmed by other experiments \cite{Jpsif0others,CDFf0life}. $R$ is observed to be very close to the original estimate. 
Recently LHCb updated the original result with 378 pb$^{-1}$ pb \cite{Conf-beta_s}. Fig.~\ref{Jpsif0}(a) shows the $J/\psi\pi^+\pi^-$ candidate mass distribution for events with $\pi^+\pi^-$ masses within $\pm$90 MeV of 980 MeV, while Fig.~\ref{Jpsif0}(b) shows the $\pi^+\pi^-$ mass distribution for events within $\pm$20 MeV of the $\Bs$ candidate mass peak. 
There is sharp peak at the $f_0(980)$ mass and an excess of events at larger masses. Measurement of the angular distributions confirms the spin-0 nature of the events in the $f_0$ peak. The higher mass region between 1200 and 1600 MeV is a mixture of spin-0 and spin-2. 

\begin{figure}[ht]
\centering
\includegraphics[width=155mm]{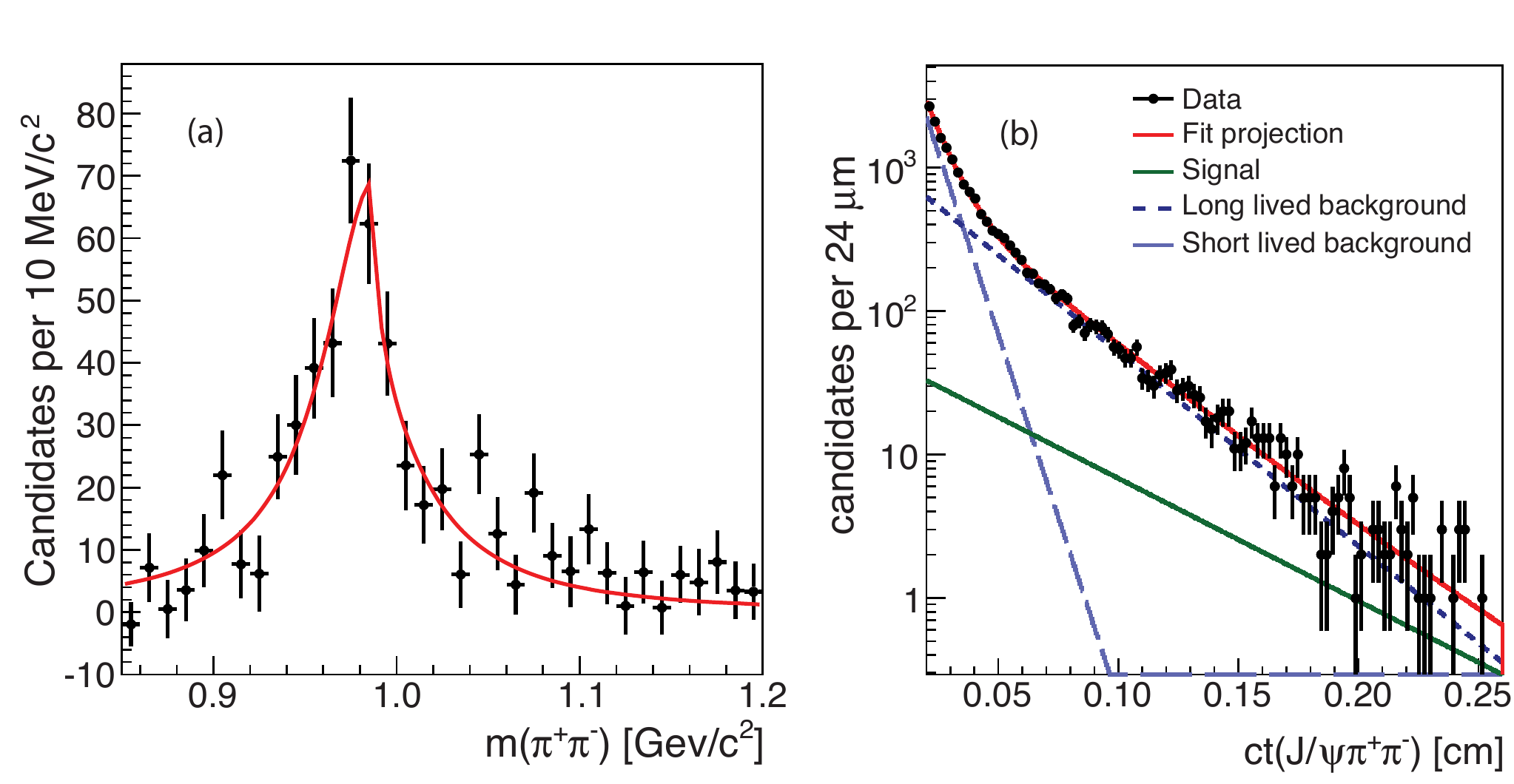}
%\vspace{-4.2cm}
\caption{$\Bs\to J/\psi f_0(980)$ data from CDF. (a) The dipion invariant mass distribution after
sideband subtraction with
fit projection overlaid. The fit uses a Flatt\'e distribution
with all parameters
floating. (b)  Decay time distribution
with fit projection overlaid.} \label{CDF-Jpsif0}
\end{figure}

The CDF collaboration measures the lifetime of this final state assuming small CP violation. The average $\Bs$ lifetime is 1.43$\pm$0.04 ps \cite{PDG}. The lifetime for this CP odd state is predicted to be larger. CDF finds a lifetime of $1.70^{+0.12}_{-0.11}\pm0.03$ ps. The fits to their data are shown in Fig.~\ref{CDF-Jpsif0}.

\begin{figure}[htb]
\centering
\includegraphics[width=165mm]{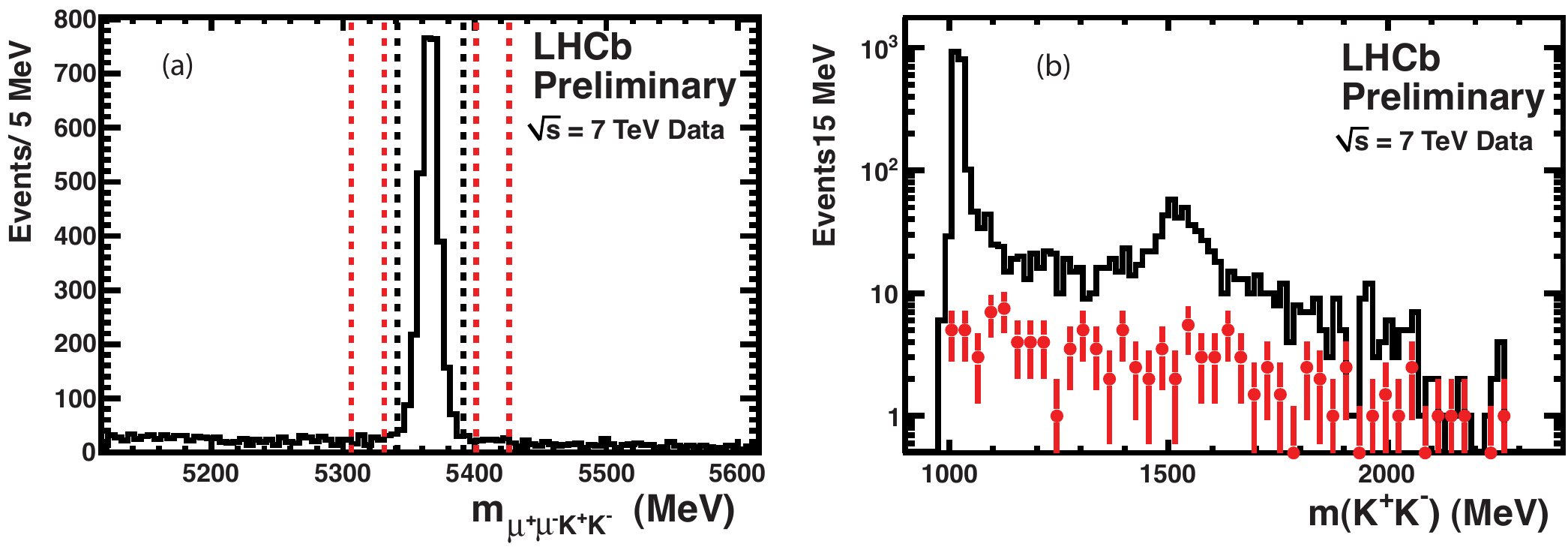}
%\vspace{-4.2cm}
\caption{(a) The invariant mass of $J/\psi K^+K^-$ combinations for the entire allowed $K^+K^-$ mass range. The vertical
lines indicate the signal and sideband regions. (b) The invariant mass of $K^+K^-$ combinations when the $J/\psi K^+K^-$ mass is required to be with $\pm$25 MeV of the $B_s$ mass. The histogram shows the data in the signal region while the points (red) show the sidebands.} \label{JpsiKK}
\end{figure}

LHCb now provides the first look at the entire $K^+K^-$ mass spectrum in $\Bs\to J/\psi K^+K^-$. The $\Bs$ candidate mass spectrum is shown in Fig.~\ref{JpsiKK}(a). Selecting the events separately in the signal and background regions the $K^+K^-$ mass spectrum is plotted in Fig.~\ref{JpsiKK}(b). 

Backgrounds across the entire mass region are small. There is a large peak at the $\phi$ mass, no surprise, but there is also a significant peak at 1525 MeV that when fit to a Breit-Wigner shape gives a mass of 1525$\pm$4 MeV and a width of $90^{+16}_{-14}$ MeV, where the uncertainties are statistical only.  A fit to the decay angular distributions shows consistency with a spin-2 resonance. Thus LHCb takes this as the known spin-2 $f'_2(1525)$ resonance. 
The effective rate is
\begin{eqnarray}
R^{f'_2}_{\rm effective}&\equiv&\frac{{\cal{B}}(B_s^0\to J/\psi f'_2(1525),~f'_2(1525)\to K^+K^-)}
{{\cal{B}}(B_s^0\to J/\psi \phi,~\phi\to K^+K^-)}=(19.4\pm1.8\pm 1.1)\% \nonumber\\ 
& &{\rm~for~}\left|m(K^+K^-) - 1525~{\rm MeV}\right|<125{\rm~MeV}.
\end{eqnarray}
These events, in principle, could also be used to measure CP violation.

\subsection{\boldmath The $X(4140)$}

\begin{figure}[htb]
\centering
\hspace*{1cm}\includegraphics[width=165mm]{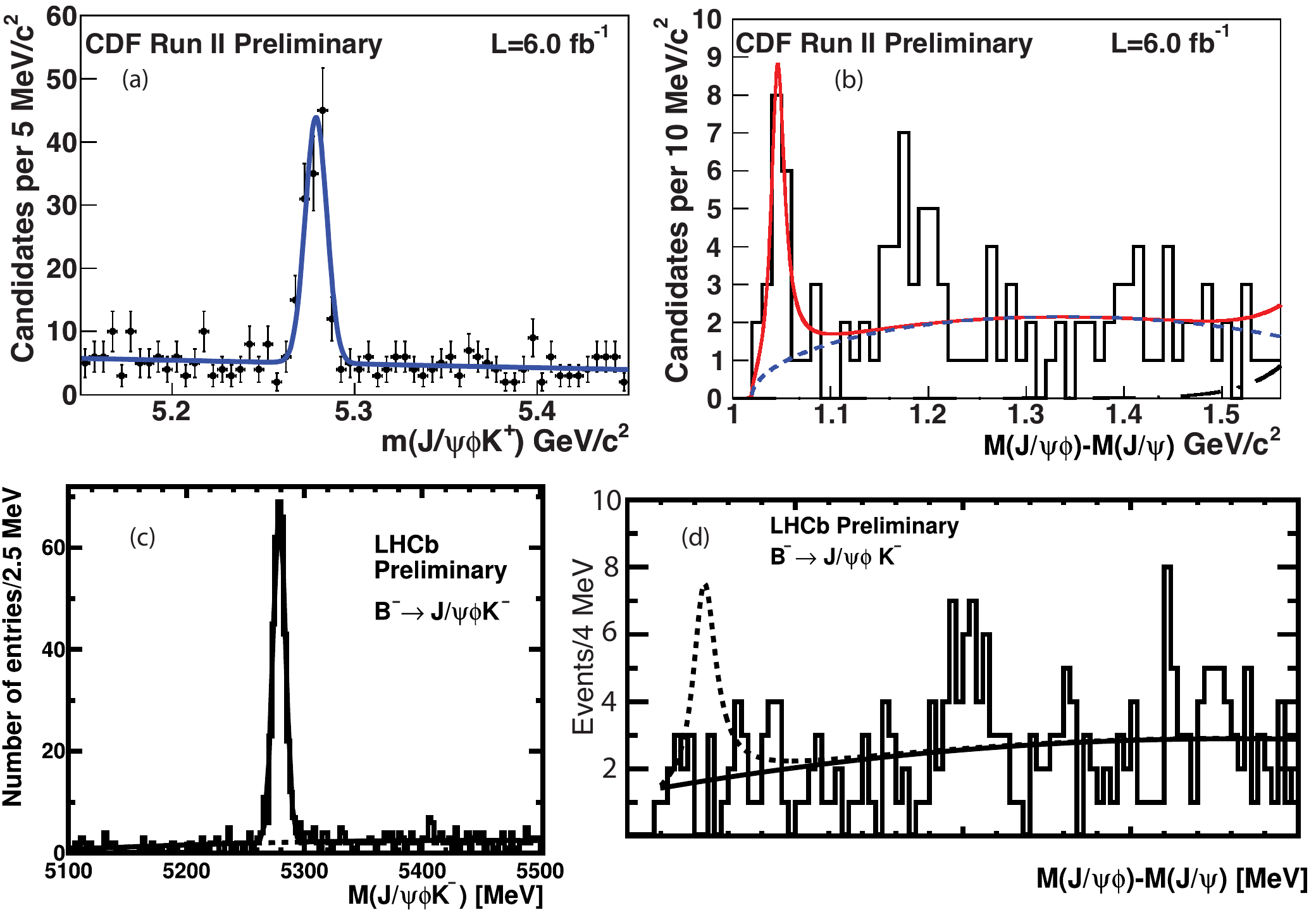}
%\vspace{-4.2cm}
\caption{(a) The $J/\psi\phi K^-$ invariant mass spectrum from (a) CDF and (c) LHCb. The $J/\psi\phi-J/\psi$ mass difference for (b) CDF and (d) LHCb. The dashed curves shows the predicted rate based on the CDF result.} \label{X4140-all}
\end{figure}

CDF has reported  not only the discovery of the $B^-\to J/\psi \phi K^-$ final state 
but also observation of a peak consistent with their mass resolution that decays into $J/\psi\phi$ \cite{CDFX}. The CDF data are shown in Fig.~\ref{X4140-all}(a) and (b). Existence of such a an exotic state could change current ideas on meson spectroscopy. The LHCb collaboration also has looked at these decays using 376 pb$^{-1}$, but unfortunately not seen such evidence \cite{LHCb4140}. The LHCb data are shown in Fig.~\ref{X4140-all}(c) and (d). LHCb has approximately 3 times the number of such $B^-$ decays. The dashed curve in (d) indicates the expected signal if the rate was that seen by CDF.

\subsection{\boldmath New $b$-baryons and Decays}

\begin{figure}[htb]
\centering
\includegraphics[width=155mm]{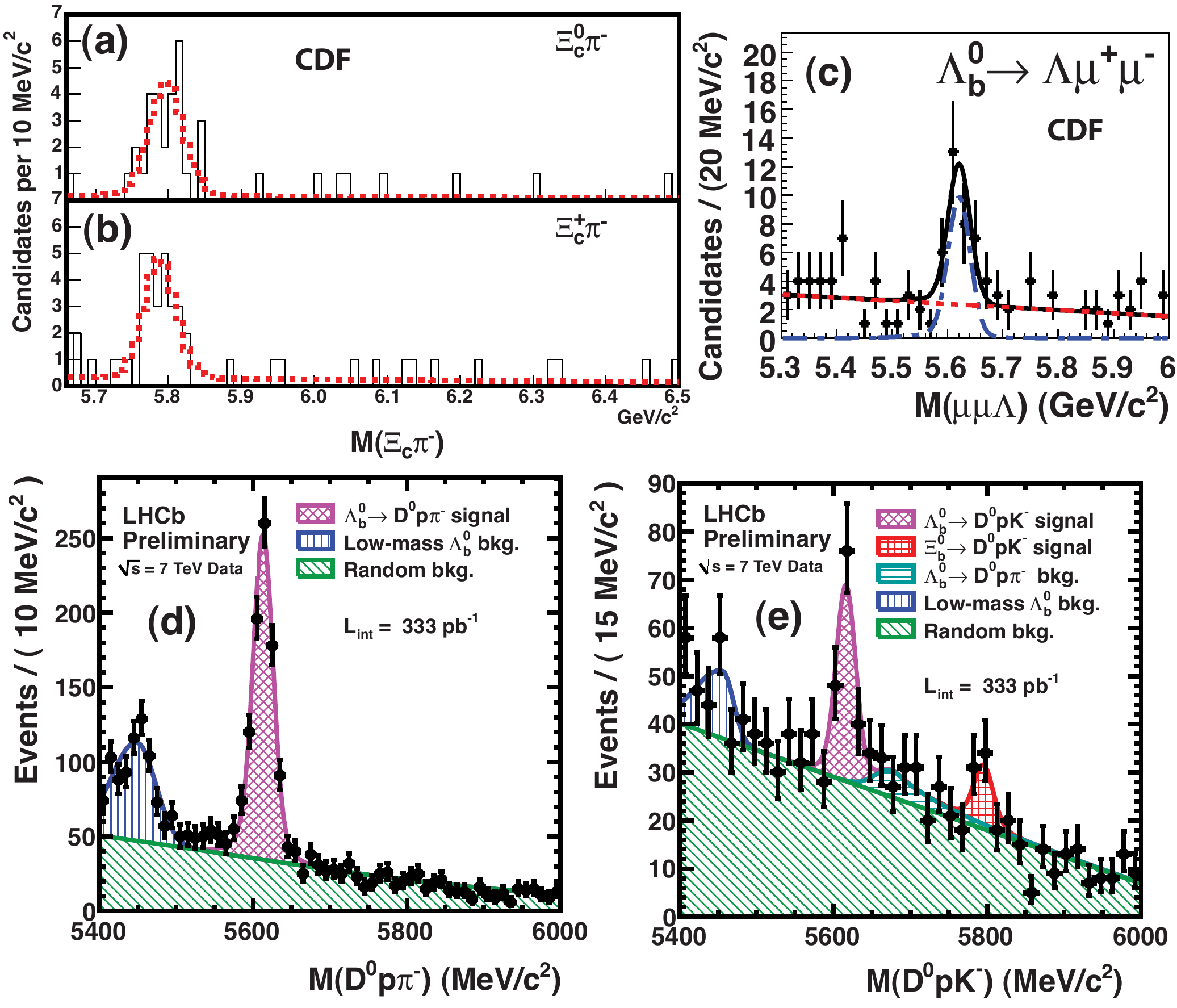}
%\vspace{-4.2cm}
\caption{Candidate mass plots. (a) CDF $\Xi_b$ for the decays (a) $\Xi_c^0\pi^-$ and (b) $\Xi_c^+\pi^-$. (c) CDF $\Lambda_b^0$ for the decay $\Lambda\mu^+\mu^-$. LHCb  for the $\Lambda_b^0$ decays (d) $D^0p\pi^-$
and (e) $D^0p K^-$; the peak at the $\Xi_b^0$ mass is not statistically significant.} \label{b-baryons}
\end{figure}

Spectroscopy studies of $b$-baryons has recently produced new results. These are shown in Fig.~\ref{b-baryons}. CDF has discovered the $\Xi_b^0$ baryon via its decay into $\Xi_c^+\pi^-$ \cite{CDF-xi0}. CDF has also observed the rare process $\Lambda_b^0\to \Lambda\mu^+\mu^-$ \cite{CDF-Lmumu}, similar to the previously discussed $B^-\to \overline{K}^{*0}\mu^+\mu^-$. LHCb has made the first observation of the process $\Lambda_b^0\to D^0 p K^-$ which could provide an alternate way of measuring the CP violating angle $\gamma$ \cite{LHCb-Lb}.

\subsection{Search for Majorana Neutrinos}
\begin{figure}[htb]
\centering
\includegraphics[width=135mm]{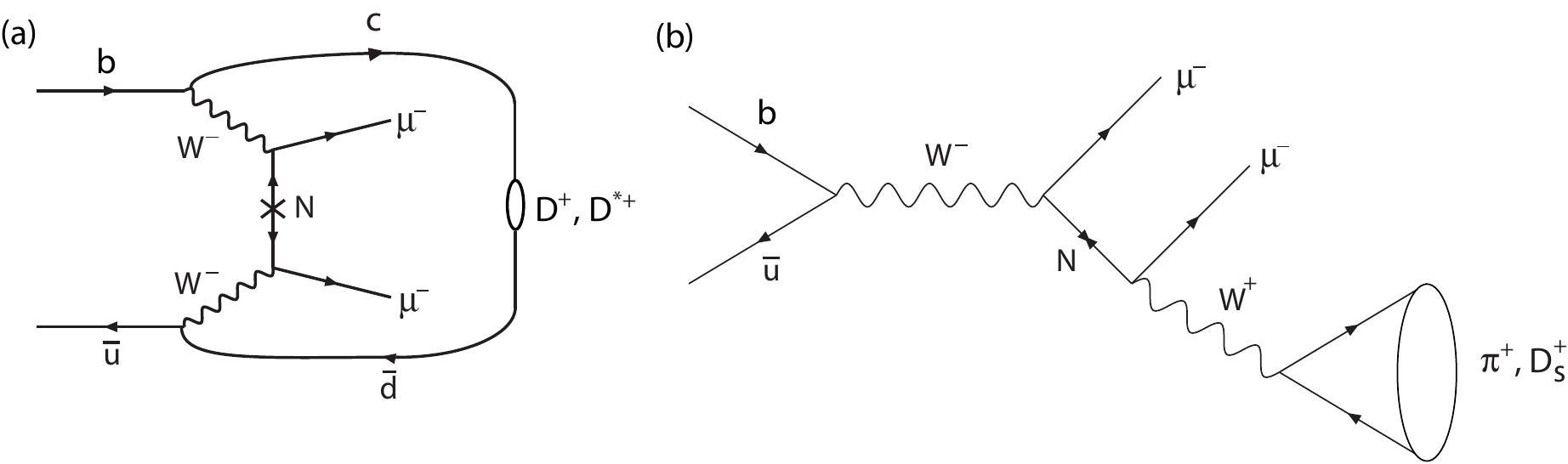}
%\vspace{-4.2cm}
\caption{Diagrams leading to like-sign dilepton events in $B^-$ decays. (a) From a virtual Majorana neutrino. (b) From a directly produced Majorana neutrino.} \label{B-Majorana4}
\end{figure}

Majorana neutrinos are distinct from SM fermions, as they are their own anti-particles. Discovery of such states would be revolutionary. It is possible to look for these objects in heavy flavor decays \cite{Han}. 
Two searches have been recently performed in $B$ meson decays. The process shown in Fig.~\ref{B-Majorana4}(a) is analogous to neutrinoless double-$\beta$ decay in nuclei. It was searched for by Belle who found upper limits on the branching fractions for $B^-\to D^+e^+e^+,$  $D^+e^+\mu^+$ and  $D^+\mu^+\mu^+$ of $2.6\times 10^{-6}$, $1.8\times 10^{-6}$, and $1.0\times 10^{-6}$, respectively at 90\% cl. This process probes all neutrino masses as the neutrino is virtual \cite{Belle-Majorana}.

\begin{figure}[htb]
\centering
\includegraphics[width=80mm]{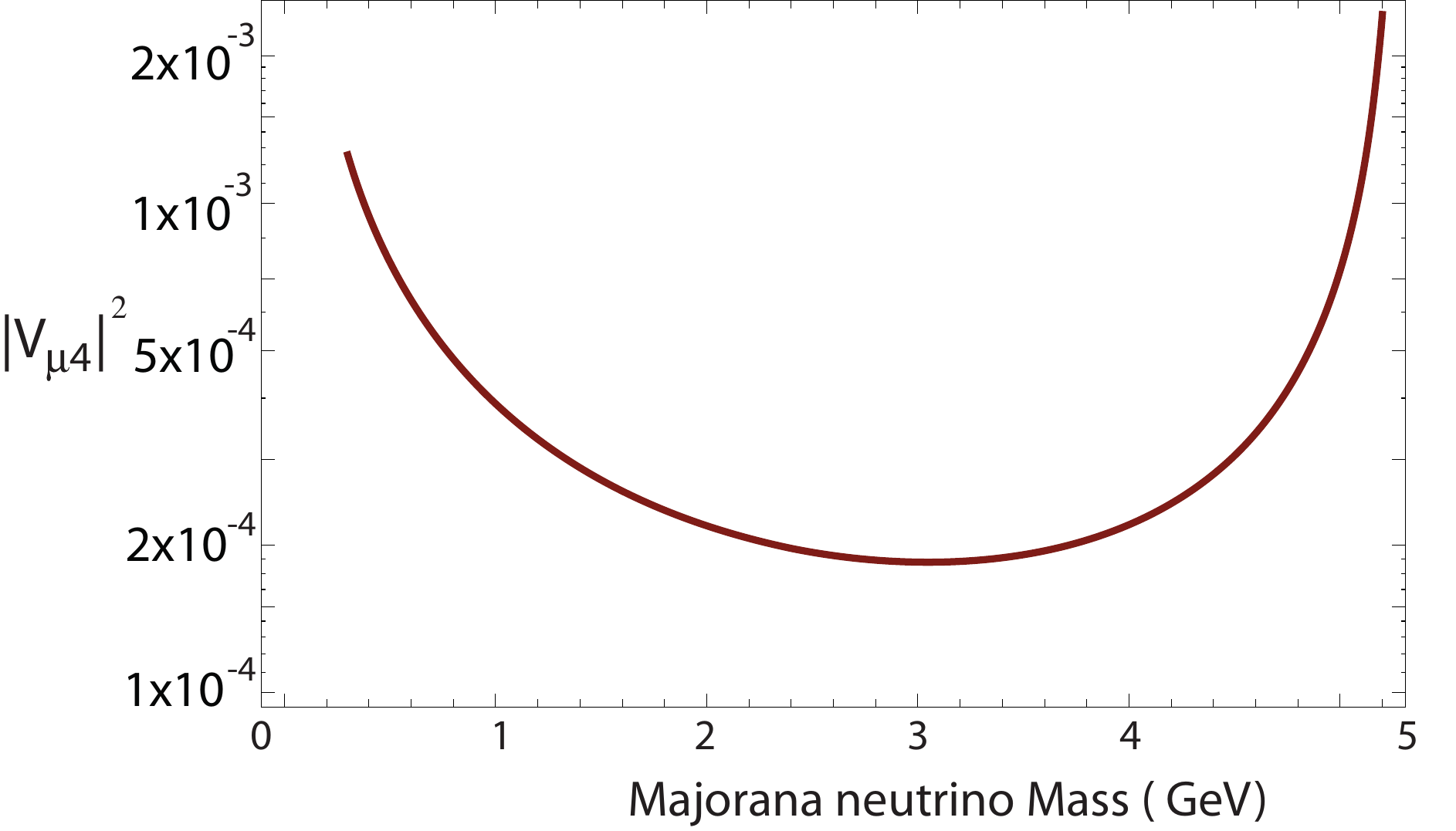}
%\vspace{-4.2cm}
\caption{Upper limits as a function of Majorana neutrino mass derived from the upper limit on ${\cal{B}}(B^-\to\pi^+\mu^-\mu^-)$ from LHCb.} \label{Majorana-pmm-ul}
\end{figure}

Another search was done by LHCb for $B^-\to\pi^+\mu^-\mu^-$ using the process shown in 
Fig.~\ref{B-Majorana4}(b).  They found an upper limit of  $<4.5\times 10^{-8}$ at 90\% cl. This limit can be translated to an upper limit on the neutrino coupling $\left|V_{\mu 4}\right|^2$ as a function of neutrino mass, using the relations in \cite{Han}. This is shown in Fig.~\ref{Majorana-pmm-ul}. However, this is only an approximate relationship as the experimental efficiency does vary with neutrino mass and this has not been taken into account.

There have also been searches for higher mass objects decaying into like-sign dileptons that could arise, for example, from real $W^-$ decays into Majorana neutrinos \cite{CDF-Mn,CMS-Mn, Atlas-Mn}. 

\subsection{Lepton Flavor Violation}
NP can also be seen by observing lepton decays that would violate lepton flavor conservation. $\tau^-$ decays to $\ell  h h$, $\Lambda h$, $\overline{\Lambda}h$, $\mu\gamma$ and $\mu\mu\mu$ have been searched for and limits approaching the $10^{-8}$ level have been set \cite{Mori,Miya}. 

\begin{figure}[htb]
\centering
\includegraphics[width=155mm]{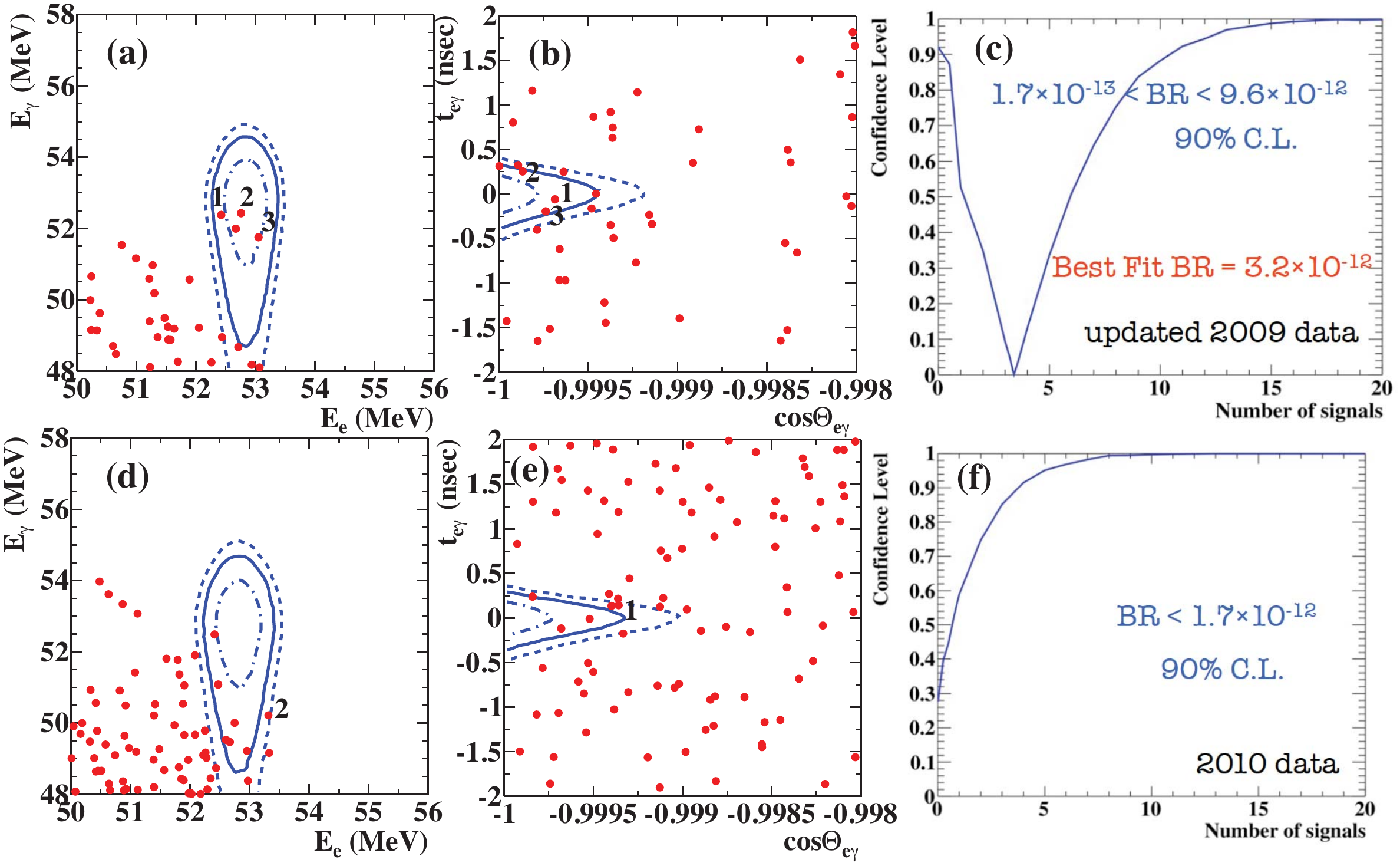}
%\vspace{-4.2cm}
\caption{MEG data for 2009 in (a-c) and 2010 (d-f). (a) and (d) show the $e^+$ energy verus the $\gamma$ energy, (b) and (e) the time difference between the $e^+$ and $\gamma$ detection versus the angle between them, and (c) and (f) show the assigned limits.} \label{MEG}
\end{figure}

The most impressive limits on a rare lepton violating process has been set by the MEG experiment for the decay $\mu^+\to e^+\gamma$. The experiment uses stopped $\mu^+$ and then they look for an $e^+$ and $\gamma$, correlated in time, back to back in angle, and each with an energy half of the $\mu^+$ mass. Their data for 2009 are shown in Fig.~\ref{MEG} \cite{Mori,MEG-paper}. A small signal is indicated in the 2009 data which led to the setting of a two sided limit. The 2010 data, however, don't show this feature and only an upper limit of $1.7\times 10^{-12}$ is set at 90\% cl. Combining both data samples, MEG sets a limit of $2.4\times 10^{-12}$. This is a very impressive limit indeed.

\section{Conclusions}

Heavy Flavor experiments have begun searching for and limiting New Physics. New results are expected soon, especially in $b$ and $c$ decays at the LHC with the 2011 data and beyond. LHCb, the first experiment designed to measure $b$ and $c$ quark decays at a hadron collider will contribute greatly in the near future, with important contributions from CMS and ATLAS.  The experiments have demonstrated their capabilities. 

New experimental initiatives are progressing. BES III has begun to take data at the $\psi(3770)$ resonance \cite{BESIII} and should have results out soon that will surpass the work of CLEO-c. Super B factories have been approved at KEK and in Italy \cite{SuperB}. An LHCb upgrade is being planned that will allow data to be collected at 5 times the current rate and will allow for a doubling of hadronic $B$ decay trigger efficiencies \cite{UpgradeLOI}.

Heavy Flavor Physics is now very sensitive to physics beyond the Standard Model. We shall see if the the first NP discoveries come from this sector or high energy searches. In either case, the study of heavy flavors will serve to distinguish the type of new physics and play a major role in categorizing it.

\begin{acknowledgments}
I thank the U. S. National Science Foundation for support. Marina Artuso, Andrei Buras, Tao Han, Mattias Neubert, Wenbin Qian and Liming Zhang are thanked for interesting discussions. I express my gratitude to Meenakshi Narain, Dave Cutts and Ulrich Heintz for the excellent organization of the meeting that made it most enjoyable. 

\end{acknowledgments}

\bigskip % extra skip inserted

% If you have acknowledgments, this puts in the proper section head.
%\bigskip % extra skip inserted
%%%%%%%%%%%%%%%%%%%%%%%%%%%%%%%%%%

% Create the reference section using BibTeX:
%\bibliography{basename of .bib file}

\begin{thebibliography}{99}  

\bibitem{reviews}
See previous reviews:  M. Artuso, E. Barberio, and S. Stone, PMC Physics A, 3:3 (2009)
arXiv:0902.3743 [hep-ph];  Y. Grossman, arXiv:1006.3534 [hep-ph]; O. Gedalia and G. Perez,  arXiv:1005.3106 [hep-ph]; G. Isidori, arXiv:1001.3431 [hep-ph]; A. J. Buras, ÒFlavour Theory: 2009,Ó arXiv:0910.1032 [hep-ph].
    
    
\bibitem{baryo}
G. R. Farrar and M. E. Shaposhnikov, Phys. Rev. Lett. {\bf 70}, 2833 (1993) [Erratum-ibid. {\bf 71},
210 (1993)] [arXiv:hep-ph/9305274]; Phys. Rev. D {\bf 50}, 774 (1994) [arXiv:hep-ph/9305275];
P. Huet and E. Sather, Phys. Rev. D {\bf 51}, 379 (1995) [arXiv:hep-ph/9404302]; M. B. Gavela,
P. Hernandez, J. Orloff, O. Pene and C. Quimbay, Nucl. Phys. B {\bf 430}, 382 (1994)
[arXiv:hep-ph/9406289].

\bibitem{dark}
M. S. Turner, [arXiv:astro-ph/9912211].
    
\bibitem{hier}
A. Masiero, S.K. Vempati, and O. Vives, [hep-ph/arXiv:0711.2903].  

\bibitem{INP}
G. Isidori, Y. Nir and G. Perez, arXiv:1002.0900 [hep-ph]; M. Neubert talk at EPS 2011,
Grenoble, France, July 2011.

\bibitem{MFV}
G. D'Ambrosio, G. Giudice, G. Isidori, and A. Strumia, Nucl. Phys. B {\bf 645}, 155 (2002)  [arXiv:hep-ph/0207036]; 
A. L. Kagan, G. Perez, T. Volansky, and J. Zupan,
Phys. Rev. D {\bf 80}, 076002 (2009) arXiv:0903.1794 [hep-ph].
    
\bibitem{Misiak}
M. Misiak \etal,~arXiv:hep-ph/0609232.
See also A. Buras \etal, arXiv:1105.5146 [hep-ph].

\bibitem{Wolf}
L. Wolfenstein, Phys. Rev. Lett. {\bf 51}, 1945 (1983).

\bibitem{PDG}
K. Nakamura \etal~(Particle Data Group), J. Phys. G {\bf 37}, 075021 (2010). 

\bibitem{CKMfitter}
J. Charles \etal~(CKMfitter Group), Eur. Phys. J. {\bf C41}, 1 (2005) [hep-ph/0406184], updated results and plots available at: http://ckmfitter.in2p3.fr~.

\bibitem{Soni}
 E. Lunghi and A. Soni, arXiv:1104.2117 [hep-ph].

\bibitem{HQE}
W.-Y. Wang, Y.-L. Wu, and F. Ye, J. Phys. {\bf G38}, 04500, (2011) 
 arXiv:1004.5444 [hep-ph] and references contained therein.
 
 \bibitem{Kowa}
 R. Kowalewski, ``$|V_{ub}|$ and $|V_{cb}|$ from Semileptonic $B$ decays and $B\to\tau\nu$," presented
 at Beauty 2011, Israel, May, 2011.
 Ê	
\bibitem{Ur}
 P. Urquijo, ``Exclusive (Semi) Leptonic Decays at Belle," presented at EPS, Grenoble, France, July, 2011.
 
\bibitem{UT}
M. Bona, ``Standard Model updates and new physics analysis with the Unitarity Triangle fit, presented at EPS, Grenoble, France, July, 2011.

\bibitem{Criv}
A. Crivellin, Phys. Rev. D{\bf 81}, 031301, (2010) arXiv:0907.2461 [hep-ph].

\bibitem{Buras-Vub}
A. J. Buras, K. Gemmler,  and G. Isidori,  arXiv:1007.1993 [hep-ph].

\bibitem{Ksmm-old}
B. Aubert \etal~(BABAR Collaboration), Phys. Rev. D {\bf 79}, 031102 (2009) arXiv:0804.4412 [hep-ex];
J.-T. Wei \etal~(Belle Collaboration, Phys. Rev. Lett. {\bf 103},  171801 (2009) arXiv:0904.0770 [hep-ex];
T. Aaltonen \etal~(CDF Collaboration),  Phys. Rev. Lett. {\bf 106}, 161801 (2011), arXiv:1101.1028 [hep-ex].

\bibitem{Ksmm-CDF}
T. Aaltonen \etal~(CDF Collaboration), arXiv:1108.0695 [hep-ex].

\bibitem{Ksmm-LHCb}
The LHCb Collaboration, LHCb-CONF-2011-038.

\bibitem{BHD}
C. Bobeth, G. Hiller, and D. van Dyk, arXiv:1105.0376 [hep-ph].

\bibitem{Fleischer}
R. Fleischer, N. Serra, and N. Tuning, Phys. Rev. D {\bf 82}, 034038 (2010) 
arXiv:1004.3982 [hep-ph].
    
\bibitem{had-frac}
R. Aaij \etal~(LHCb Collaboration), arXiv:1106.4435 [hep-ex].

\bibitem{semi-fracs}
The LHCb Collaboration, LHCb-CONF-2011-028.

\bibitem{combo-fracs}
The LHCb Collaboration, LHCb-CONF-2011-034.

     
\bibitem{Buras-mm}
A. Buras, arXiv:1012.1447 [hep-ph].

\bibitem{Thybmumu}
S. R. Choudhury and N. Gaur, Phys. Lett. {\bf B 451}, 86 (1999); K.S. Babu and C.F. Kolda, Phys. Rev. Lett. {\bf 84}, 228 (2000); M. Carena, A. Menon, and C.E.M. Wagner, Phys. Rev. D{\bf 79}, 075025 (2009); 
 arXiv:0812.3594 [hep-ph].


\bibitem{Bmumu-CDF}
T. Aaltonen \etal~(CDF Collaboration), arXiv:1107.2304 [hep-ex].

\bibitem{Kuhr}
T. Kuhr, ``Updated Search for $B_s/B_d\to\mu^+\mu^-$ at CDF," presented at EPS, Grenoble, France, July, 2011.

\bibitem{LHCb-Bmm}
The LHCb Collaboration, LHCb-CONF-2011-037.

\bibitem{CMS-Bmm}
The CMS Collaboration, CERN-PH-EP-2011-120.

\bibitem{combo}
The CMS and LHCb Collaborations, CMS-PAS-BPH-11-019; LHCb-CONF-2011-047.

\bibitem{Nonewphys}
A. G. Akeroyd, F. Mahmoudi and D. M. Santos, arXiV:1108.3018 [hep-ph], and references contained therein;
M. Farina \etal, arXiv:1104.3572 [hep-ph]; E. Golowich \etal, Phys. Rev. D{\bf 83} 114017 (2011) arXiv:1102.0009 [hep-ph];
E. Golowich  arXiv:1109.2466 [hep-ph] (update of previous paper);
O. Buchmueller \etal, Eur. Phys. J {\bf C64}, 391 (2009), arXiv:0907.5568 [hep-ph].

\bibitem {Stone-Zhang}
S. Stone and L. Zhang, Phys. Rev. D{\bf 79} (2009) 074024 
 [arXiv:0812.2832].
 
 \bibitem{Jpsif01st}
R. Aaij \etal~(LHCb Collaboration),  Phys. Lett. {\bf B} 698 (2011) 115 arXiv:1102.0206 [hep-ex].  

\bibitem{Jpsif0others}
J. Li \etal~(Belle Collaboration), Phys. Rev. Lett. {\bf 106} (2011) 121802 arXiv:1102.2759 [hep-ex];
D0 Collaboration, D0 Note 6152-CONF http://www-d0.fnal.gov/Run2Physics/WWW/results/prelim/B/B62/B62.pdf (2011).

%\bibitem{ANA-2011-001}
%R. Aaij \etal, LHCb-ANA-2011-00,
%http://cdsweb.cern.ch/record/1308189/files/LHCb-ANA-2011-001.pdf .

\bibitem{CDFf0life}
T. Aaltonen \etal,~(CDF Collaboration), arXiv:1106.3682 [hep-ex].

\bibitem{Conf-beta_s}
The LHCb Collaboration, LHCb-CONF-2011-051.

\bibitem{Conf-pipiKK}
The LHCb Collaboration, LHCb-CONF-2011-035.

\bibitem{CDFX}
T. Aaltonen \etal~(CDF),  arXiv:1101.6058 [hep-ex]; see also T. Aaltonen \etal~(CDF), Phys. Rev. Lett. {\bf 102} (2009) 242002, arXiv:0903.2229 [hep-ex].

\bibitem{LHCb4140}
The LHCb Collaboration, LHCb-CONF-2011-045.

\bibitem{CDF-xi0}
T. Aaltonen \etal~(CDF Collaboration), Phys. Rev. Lett. {\bf 107}, 102001 (2011) arXiv:1107.4015v1 [hep-ex].

\bibitem{CDF-Lmumu}
T. Aaltonen \etal~(CDF Collaboration), arXiv:1107.3753v1 [hep-ex].

\bibitem{LHCb-Lb}
The LHCb Collaboration, LHCb-CONF-2011-036.

\bibitem{Han}
A. Atre, T. Han, S. Pascoli, and B. Zhang, JHEP {\bf 0905}, 030 (2009) arXiv:0901.3589 [hep-ph]; J-M. Zhang, G-L. Wang, arXiv:1003.5570 [hep-ph]; N. Quintero, G. L. Castro and D. Delepine, arXiv:1108.6009 [hep-ph].

\bibitem{Belle-Majorana}
O. Seon \etal~(Belle Collaboration), arXiv:1107.0642 [hep-ph].

\bibitem{CDF-Mn}
A. Abulencia et. al, Phys. Rev. Lett. {\bf 98}, 221803 (2007).

\bibitem{CMS-Mn}
The CMS Collaboration, arXiv:1104.3168 [hep-ex].

\bibitem{Atlas-Mn}
The ATLAS Collaboration, arXiv:1108.0366 [hep-ex].

\bibitem{Mori}
T. Mori, talk at EPS 2011, Grenoble, France, July 2011.

\bibitem{Miya}
Y. Miyazaki, arXiv:1109.2377 [hep-ex].

\bibitem{MEG-paper}
J. Adam \etal~(MEG Collaboration), arXiv:1107.5547v3 [hep-ex].

\bibitem{BESIII}
J. Messchendorp, arXiv:1108.3047v1 [hep-ex].

\bibitem{SuperB}
P. Krizan, arXiv:1103.1209 [hep-ex].

\bibitem{UpgradeLOI}
The LHCb Collaboration, ``Letter of Intent for the LHCb Upgrade," March 2011,
http://cdsweb.cern.ch/record/1333091~.       

%\bibitem{templates-ref} http://www.slac.stanford.edu/econf/editors/eprint-template/instructions.html

\end{thebibliography}

\end{document}